\documentclass{amsart}

\usepackage{catchfile}
\usepackage{cleveref}        
\usepackage[foot]{amsaddr}   
\usepackage{fullpage}        
\usepackage{mathtools}       
\usepackage{todonotes}       
\usepackage[utf8]{inputenc}  
\usepackage{float}
\usepackage{placeins}

\DeclarePairedDelimiter\ket{\lvert}{\rangle}
\usepackage{pgfplots}
\usepackage{tikz}
\usetikzlibrary{quantikz}

\title{Efficient unstructured search implementation
\\
on current ion-trap quantum processors}
\author{Vladyslav Hlembotskyi}
\email{vlad@beit.tech}
\author{Rafał Burczyński}
\email{burek@beit.tech}
\author{Witold Jarnicki}
\email{witek@beit.tech}
\author{Adam Szady}
\email{adsz@beit.tech}
\author{Jan Tułowiecki}
\email{j.tulowiecki@beit.tech}

\begin{document}

\maketitle

\begin{abstract}
So far, only the results on 3 qubit spaces (both on superconducting and ion-trap realisations of quantum processors)  have beaten the classical unstructured search in the expected number of oracle calls using optimal protocols in both settings.
We present experimental results on running unstructured search in spaces defined by 4, 5 and 6 qubits on ion-trapped quantum processor. Our best circuits obtained respectively 66\%, 26\% and 6\% average probability of measuring the marked element. In the case of 4 and 5 qubit spaces we obtained fewer expected number of  oracle calls required to find a marked element than any classical approach. Viability of the theoretical result by Grover at these qubit counts is, to authors' knowledge demonstrated experimentally for the first time. Also at 6 qubits, a circuit using a single oracle call  returned a measured probability of success exceeding any possible classical approach.
These results were achieved using a variety of unstructured search algorithms in conjunction with  recent developments in reducing the number of entangling gates. The latter are currently considered to be a dominating source of errors in quantum computations. Some of these improvements have been made possible by using mid-circuit measurements. To our knowledge the latter feature is currently available only on the H0 quantum processor we run on.

\end{abstract}

\section{Introduction and motivation}

In the \emph{unstructured search problem} we are given an oracle which negates amplitudes of marked elements in a search space. We want to find any such an element out of $N$. It is easy to see that this problem cannot be classically solved in $o(N)$ queries to oracle. On the other hand, in the quantum setup the famous Grover's algorithm \cite{grover96} solves the problem in only $\mathcal{O}(\sqrt{N})$ oracle queries. This result is quite important and has been used in many quantum algorithms as a subroutine, see~\cite{D_rr_2006,Brassard_1998}.

\subsection{Prior work}

Since the invention of the Grover's algorithm \cite{grover96}, many attempts to run unstructured search succeeded in attaining better than classical expectation number of oracle queries only in \(2\)- and \(3\)-qubit spaces (see \cite{3q,Stromberg,Mandviwalla}). Recent attempts at getting quantum search results in \(4\)-qubit space in lower expected number of oracle queries came  short of classical ones (\cite{satoh2020subdivided,prackum2}). The latter numbers, achieved by custom, topology-aware circuits suggest that lower noise of most recent hardware can render the reimplementation successful in that respect.  The best results so far are summarized in \cref{tab:caption}. The results are averaged over oracles in column $3$, while the 4th one shows $p_\text{succ}$ for the worst performing oracle. We also provide the measure of the effectiveness of the hardware and implementation: $R=p_\text{succ}/p_t$, where $p_\text{succ}$ is  averaged over oracles probability of measuring a pattern matching the oracle and $p_t$ is the theoretical success probability of the algorithm. Our results for \(2\)- and \(3\)-qubit spaces have been published previously in \cite{prackum2}.

        \begin{table}[ht]
        \centering
        
        \begin{tabular}[t]{|l|r|r|r|c|r|} \hline
            System,space,(diffusor size,\# of iter.)& \#2q gates & $p_\text{succ}$ ave.& $p_\text{succ}$ worst & R \\ \hline\hline
               IBM Q Vigo 2-qubit (2,1) & 2 &95.18\%  & 93.32\%  & $0.95\pm0.02$ \\\hline
                IBM Q Vigo 3-qubit (3,1) & 32 & 66.14\% &60.45\% & $0.85\pm 0.07$ \\ \hline
               IBM Q Vigo 4-qubit (3,1)\cite{prackum2} & 24 & 24.5\% &19.6\% & $0.63\pm 0.05$ \\ \hline
        \end{tabular}

        \caption{Current best results.}
        \label{tab:caption}
        \end{table}

        The single iteration of Grover's algorithm as implemented in \cite{Stromberg} achieved the probability of measuring the single marked element $p_\text{succ}$ 
        equal to \(6.62\% \). However, this result was dominated by the result for oracle marking $\ket{0000}$ state, as the relaxation of the qubits to their ground 
        state results in preferred measurement outcome of $0000$. It is worth noting that the probability for a matching classical unstructured search is $6.25\%$.

    So far, as it can be seen from the \Cref{tab:caption}, quantum computers were not able to successfully run unstructured search in a space build on \(4\) qubits.

\subsection{Plan of the paper}

In \cref{sec:circuits} we present an overview of the attempted circuit families. For each family we present a brief overview, along with the
motivation for constructing a particular family, as well theoretical and expected advantages and drawbacks. When feasible, we will also present its formal
definition and a diagram. 

In \cref{sec:implementation}, we summarize the architecture-specific optimizations that we applied to our circuits, in order to fit them into the hardware.

In \cref{sec:methodology} we present the statistical background used for comparing the results coming from the hardware to the theoretical expectation.

In \cref{sec:results} we present the actual results of the hardware runs. We also perform a simple statistical
analysis introduced in \cref{sec:methodology}, in order to establish whether the amount of data obtained is sufficient to make binding claims about the results.

In \cref{sec:conclusions} we summarize the analyses performed and indicate further directions of development, both for the hardware and the software parts.

\section{Circuit types}\label{sec:circuits}

Our goal is to test each of the circuits mentioned with each qubit mask encoded as an oracle.
However, because of limited access to the hardware, we only used a subset of all possible qubit masks 
for each circuit.

For a given qubit mask $m$ we implement the baselane oracle implementation is the following:
\begin{itemize}
    \item Perform the $X$ gate on each qubit with corresponding position in $m$ set to 0.
    \item Perform a controlled Z on all qubits.
    \item Perform the $X$ gate on each qubit with corresponding position in $m$ set to 0.
\end{itemize}

For larger oracles we emply optimizations such as employing relative phase Toffoli gates and
applying partial uncompute techniques. For more details, see \cref{sec:implementation} and 
\cite{prackum}.

To test the Honeywell System Model H0 we used the following unstructured search algorithms with our
efficient implementations.

\subsection{Grover's algorithm}

The first and obvious candidate is the celebrated Grover's algorithm \cite{grover96}.
Performing the required number of iterations would cause the circuit to exceed the maximum depth
provided by hardware, hence we just perform a single operation, in order to gauge hardware 
performance.

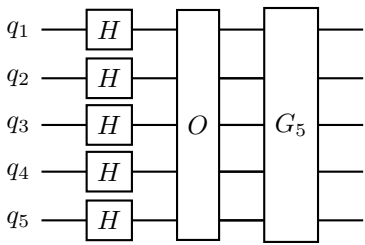
\begin{figure}[h]
    \begin{quantikz}[row sep=0.1cm, column sep=0.6cm]
        \lstick{$q_1$}     & \gate{H} & \gate[5]{O} & \gate[5]{G_5}          & \qw \\
        \lstick{$q_2$}     & \gate{H} & \ghost{O}   & \qw                  & \qw \\
        \lstick{$q_3$}     & \gate{H} & \ghost{O}   & \qw                  & \qw \\ 
        \lstick{$q_4$}     & \gate{H} & \ghost{O}   & \qw                  & \qw \\
        \lstick{$q_5$}     & \gate{H} & \ghost{O}   & \qw                  & \qw \\ 
    \end{quantikz}
    \caption{A single iteration of Grover's algorithm}
    \label{fig:grover}
\end{figure}

\subsection{Wojter}

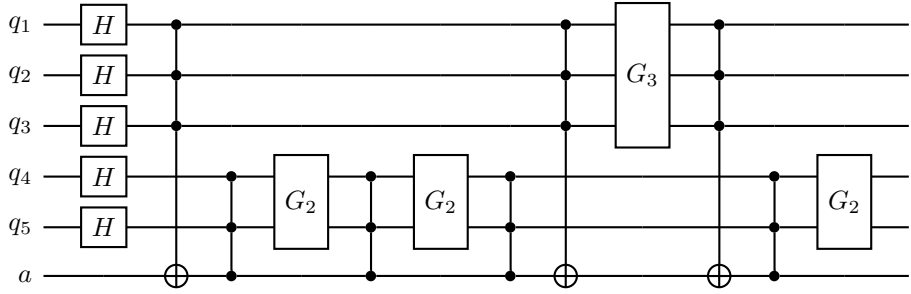
\begin{figure}[h]
    \begin{quantikz}[row sep=0.1cm, column sep=0.5cm]
        \lstick{$q_1$} & \gate{H} &  \ctrl{1} & \qw      & \qw            & \qw      & \qw            & \qw      & \ctrl{1} & \gate[3]{G_3} &  \ctrl{1} & \qw       & \qw           & \qw & \ghost{O} \\
        \lstick{$q_2$} & \gate{H} &  \ctrl{1} & \qw      & \qw            & \qw      & \qw            & \qw      & \ctrl{1} & \ghost{G_3}   &  \ctrl{1} & \qw       & \qw           & \qw & \ghost{O} \\
        \lstick{$q_3$} & \gate{H} &  \ctrl{3} & \qw      & \qw            & \qw      & \qw            & \qw      & \ctrl{3} & \ghost{G_3}   &  \ctrl{3} & \qw       & \qw           & \qw & \ghost{O} \\ 
        \lstick{$q_4$} & \gate{H} &  \qw      & \ctrl{1} & \gate[2]{G_2}  & \ctrl{1} & \gate[2]{G_2}  & \ctrl{1} & \qw      & \qw           &  \qw      & \ctrl{1}  & \gate[2]{G_2} & \qw & \ghost{O} \\
        \lstick{$q_5$} & \gate{H} &  \qw      & \ctrl{1} & \ghost{G_2}    & \ctrl{1} & \ghost{G_2}    & \ctrl{1} & \qw      & \qw           &  \qw      & \ctrl{1}  & \ghost{G_2}   & \qw & \ghost{O} \\ 
        \lstick{$a$}   & \qw      &  \targ{}  & \ctrl{}  & \qw            & \ctrl{}  & \qw            & \ctrl{}  & \targ{}  & \qw           &  \targ{}  & \ctrl{}   & \qw           & \qw & \ghost{O} \\ 
    \end{quantikz}
    \caption{$W_2$ for $\overline{k} = (3, 2)$}
    \label{fig:wojter}
\end{figure}

We run circuits \(W_n\) proposed in \cite{prackum} on \(4\) and \(5\) qubit search spaces, those circuits combined with amplitude amplification asymptotically constitute optimal quantum search algorithms while allowing for minimal number of additional gates. The circuit description is presented in \cref{fig:wojter}. In \cref{sec:implementation} we show how to effectively utilize simplified \(CCX\) and \(CCCX\) gates to reduce the number of 2-qubit gates.

\subsection{Drzewker}
Circuit family \(D_n\) also allows for optimal quantum search algorithms as proved in \cite{prackum2} and presents a significant asymptotic improvement when it comes to the number of non-oracle gates.

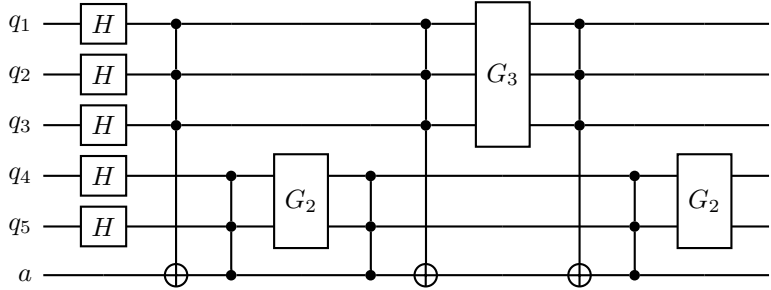
\begin{figure}[h]
    \begin{quantikz}[row sep=0.1cm, column sep=0.5cm]
        \lstick{$q_1$} & \gate{H} &  \ctrl{1} & \qw      & \qw            & \qw      &  \ctrl{1} & \gate[3]{G_3} &  \ctrl{1} & \qw       & \qw           & \qw & \ghost{O} \\
        \lstick{$q_2$} & \gate{H} &  \ctrl{1} & \qw      & \qw            & \qw      &  \ctrl{1} & \ghost{G_3}   &  \ctrl{1} & \qw       & \qw           & \qw & \ghost{O} \\
        \lstick{$q_3$} & \gate{H} &  \ctrl{3} & \qw      & \qw            & \qw      &  \ctrl{3} & \ghost{G_3}   &  \ctrl{3} & \qw       & \qw           & \qw & \ghost{O} \\ 
        \lstick{$q_4$} & \gate{H} &  \qw      & \ctrl{1} & \gate[2]{G_2}  & \ctrl{1} &  \qw      & \qw           &  \qw      & \ctrl{1}  & \gate[2]{G_2} & \qw & \ghost{O} \\
        \lstick{$q_5$} & \gate{H} &  \qw      & \ctrl{1} & \ghost{G_2}    & \ctrl{1} &  \qw      & \qw           &  \qw      & \ctrl{1}  & \ghost{G_2}   & \qw & \ghost{O} \\ 
        \lstick{$a$}   & \qw      &  \targ{}  & \ctrl{}  & \qw            & \ctrl{}  &  \targ{}  & \qw           &  \targ{}  & \ctrl{}   & \qw           & \qw & \ghost{O} \\ 
    \end{quantikz}
    \caption{$D_2$ for $\overline{k} = (3, 2)$}
    \label{fig:drzewker}
\end{figure}

\subsection{Wielomianer}

\begin{figure}[h]
    \begin{quantikz}[row sep=0.1cm, column sep=0.6cm]
        \lstick{$q_1$}     & \gate{H} & \ctrl{1}  &  \qw       & \qw           & \qw      & \qw      & \gate[2]{G_2} & \ctrl{1}    & \qw         & \qw           & \qw     & \ghost{O} \\
        \lstick{$q_2$}     & \gate{H} & \ctrl{1}  &  \qw       & \qw           & \qw      & \qw      & \ghost{G_2}   & \ctrl{1}    & \qw         & \qw           & \qw     & \ghost{O} \\
        \lstick{$a$}       & \qw      & \targ{}   &  \ctrl{}   & \qw           & \ctrl{1} & \qw      & \qw           & \targ{}     & \ctrl{}     & \qw           & \qw     & \ghost{O} \\
        \lstick{$q_3$}     & \gate{H} & \qw       &  \ctrl{-1} & \gate[2]{G_2} & \ctrl{1} & \qw      & \qw           & \qw         & \ctrl{-1}   & \gate[2]{G_2} & \qw     & \ghost{O} \\
        \lstick{$q_4$}     & \gate{H} & \qw       &  \ctrl{-1} & \ghost{G_2}   & \ctrl{1} & \qw      & \qw           & \qw         & \ctrl{-1}   & \ghost{G_2}   & \qw     & \ghost{O} \\
        \lstick{$\ket{0}$} & \qw      & \qw       &  \qw       & \qw           & \targ{}  & \meter{} & \octrl{} \vcw{-4}   & \octrl{} \vcw{-3} & \octrl{} \vcw{-1} & \octrl{} \vcw{-1}   & \qw     & \ghost{O} \\
    \end{quantikz}
    \caption{$P_{4,3}$ algorithm}
    \label{fig:wielomianer_43}
\end{figure}
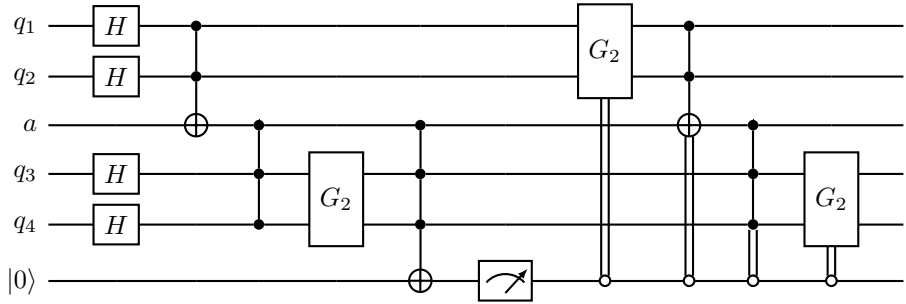

We introduce a family of machine-generated circuits and discuss some of its properties. We execute an algorithm $P_{4,3}$ from this family, presented in Fig. \ref{fig:wielomianer_43}.

 \section{Implementation}\label{sec:implementation}

To implement our circuits as efficient as possible we use a few optimizations to 
expedite unstructured search. We try different algorithm for solving the unstructured 
search problems as described above. Moreover, we want to implement oracles and
diffusion operators in a way which minimizes the total number of $ZZ$ basic gates. 

We generally use the same optimization techniques described in \cite{prackum2}. We also 
use relative phase Toffoli gate \cite{adamolus} instead of Margolus gate to implement 
larger oracles. This gate can be used as a replacement when implementing the oracle and 
it requires fewer 2-qubit basic gates to implement. Fortunately, the Honeywell machine 
has a clique topology (logical all-to-all connectivity), this way it is much easier to implement various 
algorithms. 
On the other hand, during our experiments a temporary hardware issue caused reduced performance on the
$5^{th}$ and $6^{th}$ qubits with certain circuit implementations. We adapted our algorithms to best 
accommodate the machine implementation when mapping the circuits.

\section{Results analysis methodology}\label{sec:methodology}

For each circuit considered, we plot the ideal theoretical distribution, along with the
experimental results, average over all oracles used. To make the results for different 
oracles compatible for computing the average, the value $x$ on the ``pattern'' axis 
corresponds to the qbit value $x\oplus x_0$, where $x_0$ is the element marked by the 
oracle. For more details, see \cite{prackum2}.

The experiments were performed during two three-hour sessions. If similar experiments
were performed during two sessions the two plots are placed next to each other.
Feedback from the first session resulted in hardware improvements like continuous calibration, which led to better results of the second set of runs.

\newenvironment{ffigure}{\let\savecaption\caption\def\caption##1{\\##1\\}\begin{center}}%
{\end{center}\let\caption\savecaption}

\section{Experiment results}\label{sec:results}
The results are grouped according to the number of qubits the search was performed on.
\subsection{3-qubit search}

\begin{ffigure}
\begin{tikzpicture}
\begin{axis}[
    height=8.6cm,  
    width=8.4cm,
    xmin=-0.6,xmax=7.6,ymin=-0.1,ymax=660,
    legend cell align=left,
    ylabel ={count}, ylabel near ticks, yticklabel pos=left,
    xlabel={pattern},
    xlabel near ticks, xticklabel pos=bottom,
]
\addplot [thick, mark=none] coordinates {
   (-0.5,0)
   (-0.5,472.5)
   (0.5,472.5)
   (0.5,4.0)
   (1.5,4.0)
   (1.5,4.0)
   (2.5,4.0)
   (2.5,4.0)
   (3.5,4.0)
   (3.5,4.0)
   (4.5,4.0)
   (4.5,4.0)
   (5.5,4.0)
   (5.5,4.0)
   (6.5,4.0)
   (6.5,4.0)
   (7.5,4.0)
   (7.5,0)
};
\addplot [dashed, mark=none] coordinates { 
   (-0.5,0)
   (-0.5,472.5+29.16333314283537)
   (0.5,472.5+29.16333314283537)
   (0.5,4.0+2.6832815729997477)
   (1.5,4.0+2.6832815729997477)
   (1.5,4.0+2.6832815729997477)
   (2.5,4.0+2.6832815729997477)
   (2.5,4.0+2.6832815729997477)
   (3.5,4.0+2.6832815729997477)
   (3.5,4.0+2.6832815729997477)
   (4.5,4.0+2.6832815729997477)
   (4.5,4.0+2.6832815729997477)
   (5.5,4.0+2.6832815729997477)
   (5.5,4.0+2.6832815729997477)
   (6.5,4.0+2.6832815729997477)
   (6.5,4.0+2.6832815729997477)
   (7.5,4.0+2.6832815729997477)
   (7.5,0)
};
\addplot [only marks, error bars,y dir=both, y explicit] coordinates {
(0, 376.25) += (0, 9.75) -= (0, 18.25)
(1, 18.0) += (1, 3.0) -= (1, 3.0)
(2, 18.75) += (2, 7.25) -= (2, 8.75)
(3, 14.25) += (3, 5.75) -= (3, 4.25)
(4, 18.5) += (4, 3.5) -= (4, 6.5)
(5, 14.75) += (5, 7.25) -= (5, 7.75)
(6, 24.0) += (6, 6.0) -= (6, 10.0)
(7, 15.5) += (7, 2.5) -= (7, 1.5)
};
\addplot [dashed, mark=none] coordinates {
   (-0.5,0)
   (-0.5,472.5-29.16333314283537)
   (0.5,472.5-29.16333314283537)
   (0.5,4.0-2.6832815729997477)
   (1.5,4.0-2.6832815729997477)
   (1.5,4.0-2.6832815729997477)
   (2.5,4.0-2.6832815729997477)
   (2.5,4.0-2.6832815729997477)
   (3.5,4.0-2.6832815729997477)
   (3.5,4.0-2.6832815729997477)
   (4.5,4.0-2.6832815729997477)
   (4.5,4.0-2.6832815729997477)
   (5.5,4.0-2.6832815729997477)
   (5.5,4.0-2.6832815729997477)
   (6.5,4.0-2.6832815729997477)
   (6.5,4.0-2.6832815729997477)
   (7.5,4.0-2.6832815729997477)
   (7.5,0)
};
\addlegendentry{3-Grover-b}
\addlegendentry {3 RMS Error, 5$\times$500 shots}
\addlegendentry{Honeywell {\bf 75.2\%}} 
\end{axis}
\end{tikzpicture}
\begin{tikzpicture}
\begin{axis}[
    height=8.6cm,  
    width=8.4cm,
    xmin=-0.6,xmax=7.6,ymin=-0.1,ymax=660,
    legend cell align=left,
    ylabel ={count}, ylabel near ticks, yticklabel pos=right,
    xlabel={pattern},
    xlabel near ticks, xticklabel pos=bottom,
]
\addplot [thick, mark=none] coordinates {
   (-0.5,0)
   (-0.5,472.5)
   (0.5,472.5)
   (0.5,4.0)
   (1.5,4.0)
   (1.5,4.0)
   (2.5,4.0)
   (2.5,4.0)
   (3.5,4.0)
   (3.5,4.0)
   (4.5,4.0)
   (4.5,4.0)
   (5.5,4.0)
   (5.5,4.0)
   (6.5,4.0)
   (6.5,4.0)
   (7.5,4.0)
   (7.5,0)
};
\addplot [dashed, mark=none] coordinates { 
   (-0.5,0)
   (-0.5,472.5+37.6497011940334)
   (0.5,472.5+37.6497011940334)
   (0.5,4.0+3.464101615137755)
   (1.5,4.0+3.464101615137755)
   (1.5,4.0+3.464101615137755)
   (2.5,4.0+3.464101615137755)
   (2.5,4.0+3.464101615137755)
   (3.5,4.0+3.464101615137755)
   (3.5,4.0+3.464101615137755)
   (4.5,4.0+3.464101615137755)
   (4.5,4.0+3.464101615137755)
   (5.5,4.0+3.464101615137755)
   (5.5,4.0+3.464101615137755)
   (6.5,4.0+3.464101615137755)
   (6.5,4.0+3.464101615137755)
   (7.5,4.0+3.464101615137755)
   (7.5,0)
};
\addplot [only marks, error bars,y dir=both, y explicit] coordinates {
(0, 360.0) += (0, 7.0) -= (0, 10.0)
(1, 16.0) += (1, 6.0) -= (1, 7.0)
(2, 22.65) += (2, 11.350000000000001) -= (2, 8.65)
(3, 17.0) += (3, 5.0) -= (3, 4.0)
(4, 30.65) += (4, 1.35) -= (4, 0.65)
(5, 14.65) += (5, 5.35) -= (5, 5.6499999999999995)
(6, 18.35) += (6, 7.6499999999999995) -= (6, 5.35)
(7, 20.650000000000002) += (7, 4.35) -= (7, 2.65)
};
\addplot [dashed, mark=none] coordinates {
   (-0.5,0)
   (-0.5,472.5-37.6497011940334)
   (0.5,472.5-37.6497011940334)
   (0.5,4.0-3.464101615137755)
   (1.5,4.0-3.464101615137755)
   (1.5,4.0-3.464101615137755)
   (2.5,4.0-3.464101615137755)
   (2.5,4.0-3.464101615137755)
   (3.5,4.0-3.464101615137755)
   (3.5,4.0-3.464101615137755)
   (4.5,4.0-3.464101615137755)
   (4.5,4.0-3.464101615137755)
   (5.5,4.0-3.464101615137755)
   (5.5,4.0-3.464101615137755)
   (6.5,4.0-3.464101615137755)
   (6.5,4.0-3.464101615137755)
   (7.5,4.0-3.464101615137755)
   (7.5,0)
};
\addlegendentry{3-Grover-b}
\addlegendentry {3 RMS Error, 3$\times$500 shots}
\addlegendentry{Honeywell {\bf 72.0\%}} 
\end{axis}
\end{tikzpicture}
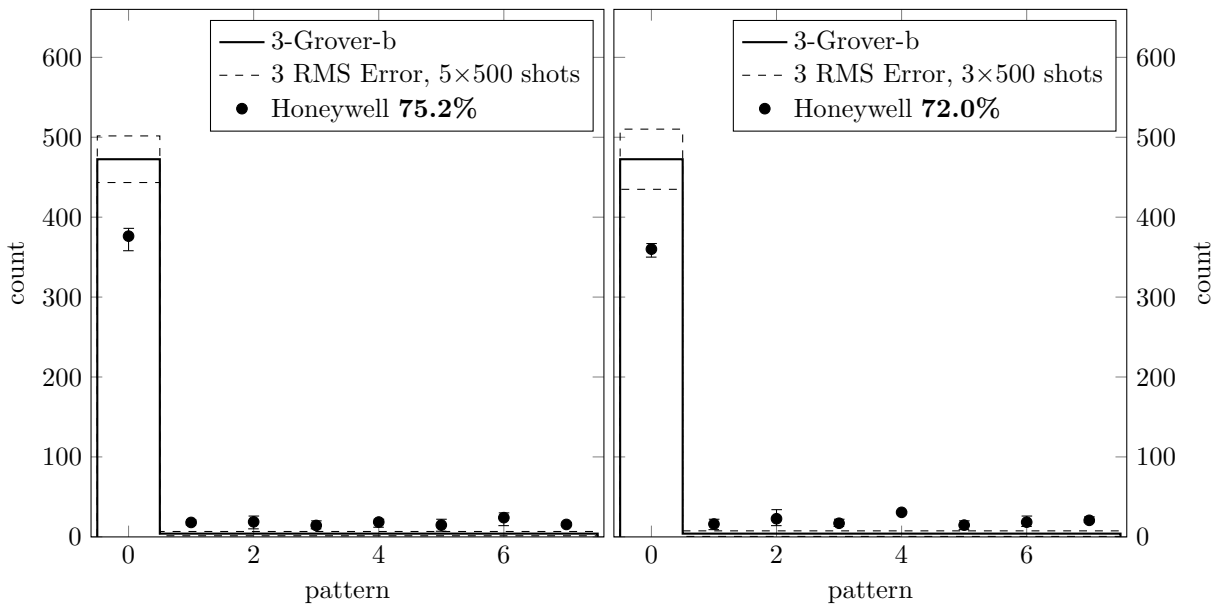
\caption{3-qubit Grover: session 1 (left), session 2 (right).}
\end{ffigure}

\begin{ffigure} 
\begin{tikzpicture}
\begin{axis}[
    height=8.6cm,  
    width=8.4cm,
    xmin=-0.6,xmax=7.6,ymin=-0.1,ymax=660,
    legend cell align=left,
    ylabel ={count}, ylabel near ticks, yticklabel pos=right,
    xlabel={pattern},
    xlabel near ticks, xticklabel pos=bottom,
]
\addplot [thick, mark=none] coordinates {
   (-0.5,0)
   (-0.5,500.0)
   (0.5,500.0)
   (0.5,0.0)
   (1.5,0.0)
   (1.5,0.0)
   (2.5,0.0)
   (2.5,0.0)
   (3.5,0.0)
   (3.5,0.0)
   (4.5,0.0)
   (4.5,0.0)
   (5.5,0.0)
   (5.5,0.0)
   (6.5,0.0)
   (6.5,0.0)
   (7.5,0.0)
   (7.5,0)
};
\addplot [dashed, mark=none] coordinates { 
   (-0.5,0)
   (-0.5,500.0+30.0)
   (0.5,500.0+30.0)
   (0.5,0.0+0.0)
   (1.5,0.0+0.0)
   (1.5,0.0+0.0)
   (2.5,0.0+0.0)
   (2.5,0.0+0.0)
   (3.5,0.0+0.0)
   (3.5,0.0+0.0)
   (4.5,0.0+0.0)
   (4.5,0.0+0.0)
   (5.5,0.0+0.0)
   (5.5,0.0+0.0)
   (6.5,0.0+0.0)
   (6.5,0.0+0.0)
   (7.5,0.0+0.0)
   (7.5,0)
};
\addplot [only marks, error bars,y dir=both, y explicit] coordinates {
(0, 318.0) += (0, 84.0) -= (0, 54.0)
(1, 107.64999999999999) += (1, 53.35) -= (1, 77.64999999999999)
(2, 15.65) += (2, 14.35) -= (2, 7.6499999999999995)
(3, 17.0) += (3, 8.0) -= (3, 13.0)
(4, 15.0) += (4, 13.0) -= (4, 7.0)
(5, 13.65) += (5, 9.350000000000001) -= (5, 13.65)
(6, 4.0) += (6, 2.0) -= (6, 2.0)
(7, 9.0) += (7, 11.0) -= (7, 9.0)
};
\addplot [dashed, mark=none] coordinates {
   (-0.5,0)
   (-0.5,500.0-30.0)
   (0.5,500.0-30.0)
   (0.5,0.0-0.0)
   (1.5,0.0-0.0)
   (1.5,0.0-0.0)
   (2.5,0.0-0.0)
   (2.5,0.0-0.0)
   (3.5,0.0-0.0)
   (3.5,0.0-0.0)
   (4.5,0.0-0.0)
   (4.5,0.0-0.0)
   (5.5,0.0-0.0)
   (5.5,0.0-0.0)
   (6.5,0.0-0.0)
   (6.5,0.0-0.0)
   (7.5,0.0-0.0)
   (7.5,0)
};
\addlegendentry{3-Wielomianer-1am-b}
\addlegendentry {3 RMS Error, 5$\times$500 shots}
\addlegendentry{Honeywell {\bf 63.6\%}} 
\end{axis}
\end{tikzpicture}
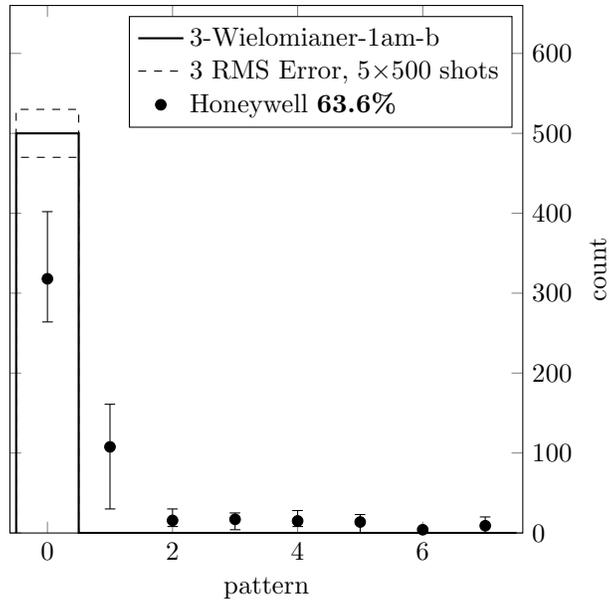
\caption{3-qubit machine-generated ``Wielomianer'' circuit with mid-circuit measurement.}
\end{ffigure}

\subsection{4-qubit search}

\begin{ffigure}
\begin{tikzpicture}
\begin{axis}[
    height=8.6cm,  
    width=8.4cm,
    xmin=-0.6,xmax=15.6,ymin=-0.1,ymax=660,
    legend cell align=left,
    ylabel ={count}, ylabel near ticks, yticklabel pos=left,
    xlabel={pattern},
    xlabel near ticks, xticklabel pos=bottom,
]
\addplot [thick, mark=none] coordinates {
   (-0.5,0)
   (-0.5,500.0)
   (0.5,500.0)
   (0.5,0.0)
   (1.5,0.0)
   (1.5,0.0)
   (2.5,0.0)
   (2.5,0.0)
   (3.5,0.0)
   (3.5,0.0)
   (4.5,0.0)
   (4.5,0.0)
   (5.5,0.0)
   (5.5,0.0)
   (6.5,0.0)
   (6.5,0.0)
   (7.5,0.0)
   (7.5,0.0)
   (8.5,0.0)
   (8.5,0.0)
   (9.5,0.0)
   (9.5,0.0)
   (10.5,0.0)
   (10.5,0.0)
   (11.5,0.0)
   (11.5,0.0)
   (12.5,0.0)
   (12.5,0.0)
   (13.5,0.0)
   (13.5,0.0)
   (14.5,0.0)
   (14.5,0.0)
   (15.5,0.0)
   (15.5,0)
};
\addplot [dashed, mark=none] coordinates { 
   (-0.5,0)
   (-0.5,500.0+38.72983346207417)
   (0.5,500.0+38.72983346207417)
   (0.5,0.0+0.0)
   (1.5,0.0+0.0)
   (1.5,0.0+0.0)
   (2.5,0.0+0.0)
   (2.5,0.0+0.0)
   (3.5,0.0+0.0)
   (3.5,0.0+0.0)
   (4.5,0.0+0.0)
   (4.5,0.0+0.0)
   (5.5,0.0+0.0)
   (5.5,0.0+0.0)
   (6.5,0.0+0.0)
   (6.5,0.0+0.0)
   (7.5,0.0+0.0)
   (7.5,0.0+0.0)
   (8.5,0.0+0.0)
   (8.5,0.0+0.0)
   (9.5,0.0+0.0)
   (9.5,0.0+0.0)
   (10.5,0.0+0.0)
   (10.5,0.0+0.0)
   (11.5,0.0+0.0)
   (11.5,0.0+0.0)
   (12.5,0.0+0.0)
   (12.5,0.0+0.0)
   (13.5,0.0+0.0)
   (13.5,0.0+0.0)
   (14.5,0.0+0.0)
   (14.5,0.0+0.0)
   (15.5,0.0+0.0)
   (15.5,0)
};
\addplot [only marks, error bars,y dir=both, y explicit] coordinates {
(0, 247.65) += (0, 136.35) -= (0, 121.64999999999999)
(1, 31.35) += (1, 18.65) -= (1, 23.349999999999998)
(2, 21.0) += (2, 22.0) -= (2, 12.0)
(3, 13.350000000000001) += (3, 14.65) -= (3, 9.350000000000001)
(4, 31.65) += (4, 12.35) -= (4, 7.6499999999999995)
(5, 23.349999999999998) += (5, 22.65) -= (5, 21.35)
(6, 14.35) += (6, 15.65) -= (6, 11.350000000000001)
(7, 6.35) += (7, 8.65) -= (7, 5.35)
(8, 18.0) += (8, 1.0) -= (8, 2.0)
(9, 21.0) += (9, 12.0) -= (9, 19.0)
(10, 10.0) += (10, 10.0) -= (10, 7.0)
(11, 8.0) += (11, 7.0) -= (11, 5.0)
(12, 8.65) += (12, 6.35) -= (12, 4.6499999999999995)
(13, 23.349999999999998) += (13, 9.65) -= (13, 15.350000000000001)
(14, 13.65) += (14, 9.350000000000001) -= (14, 4.6499999999999995)
(15, 8.35) += (15, 9.65) -= (15, 5.35)
};
\addplot [dashed, mark=none] coordinates {
   (-0.5,0)
   (-0.5,500.0-38.72983346207417)
   (0.5,500.0-38.72983346207417)
   (0.5,0.0-0.0)
   (1.5,0.0-0.0)
   (1.5,0.0-0.0)
   (2.5,0.0-0.0)
   (2.5,0.0-0.0)
   (3.5,0.0-0.0)
   (3.5,0.0-0.0)
   (4.5,0.0-0.0)
   (4.5,0.0-0.0)
   (5.5,0.0-0.0)
   (5.5,0.0-0.0)
   (6.5,0.0-0.0)
   (6.5,0.0-0.0)
   (7.5,0.0-0.0)
   (7.5,0.0-0.0)
   (8.5,0.0-0.0)
   (8.5,0.0-0.0)
   (9.5,0.0-0.0)
   (9.5,0.0-0.0)
   (10.5,0.0-0.0)
   (10.5,0.0-0.0)
   (11.5,0.0-0.0)
   (11.5,0.0-0.0)
   (12.5,0.0-0.0)
   (12.5,0.0-0.0)
   (13.5,0.0-0.0)
   (13.5,0.0-0.0)
   (14.5,0.0-0.0)
   (14.5,0.0-0.0)
   (15.5,0.0-0.0)
   (15.5,0)
};
\addlegendentry{4-Wojter-2ac-pu}
\addlegendentry {3 RMS Error, 3$\times$500 shots}
\addlegendentry{Honeywell {\bf 49.5\%}} 
\end{axis}
\end{tikzpicture}
\begin{tikzpicture}
\begin{axis}[
    height=8.6cm,  
    width=8.4cm,
    xmin=-0.6,xmax=15.6,ymin=-0.1,ymax=660,
    legend cell align=left,
    ylabel ={count}, ylabel near ticks, yticklabel pos=right,
    xlabel={pattern},
    xlabel near ticks, xticklabel pos=bottom,
]
\addplot [thick, mark=none] coordinates {
   (-0.5,0)
   (-0.5,500.0)
   (0.5,500.0)
   (0.5,0.0)
   (1.5,0.0)
   (1.5,0.0)
   (2.5,0.0)
   (2.5,0.0)
   (3.5,0.0)
   (3.5,0.0)
   (4.5,0.0)
   (4.5,0.0)
   (5.5,0.0)
   (5.5,0.0)
   (6.5,0.0)
   (6.5,0.0)
   (7.5,0.0)
   (7.5,0.0)
   (8.5,0.0)
   (8.5,0.0)
   (9.5,0.0)
   (9.5,0.0)
   (10.5,0.0)
   (10.5,0.0)
   (11.5,0.0)
   (11.5,0.0)
   (12.5,0.0)
   (12.5,0.0)
   (13.5,0.0)
   (13.5,0.0)
   (14.5,0.0)
   (14.5,0.0)
   (15.5,0.0)
   (15.5,0)
};
\addplot [dashed, mark=none] coordinates { 
   (-0.5,0)
   (-0.5,500.0+38.72983346207417)
   (0.5,500.0+38.72983346207417)
   (0.5,0.0+0.0)
   (1.5,0.0+0.0)
   (1.5,0.0+0.0)
   (2.5,0.0+0.0)
   (2.5,0.0+0.0)
   (3.5,0.0+0.0)
   (3.5,0.0+0.0)
   (4.5,0.0+0.0)
   (4.5,0.0+0.0)
   (5.5,0.0+0.0)
   (5.5,0.0+0.0)
   (6.5,0.0+0.0)
   (6.5,0.0+0.0)
   (7.5,0.0+0.0)
   (7.5,0.0+0.0)
   (8.5,0.0+0.0)
   (8.5,0.0+0.0)
   (9.5,0.0+0.0)
   (9.5,0.0+0.0)
   (10.5,0.0+0.0)
   (10.5,0.0+0.0)
   (11.5,0.0+0.0)
   (11.5,0.0+0.0)
   (12.5,0.0+0.0)
   (12.5,0.0+0.0)
   (13.5,0.0+0.0)
   (13.5,0.0+0.0)
   (14.5,0.0+0.0)
   (14.5,0.0+0.0)
   (15.5,0.0+0.0)
   (15.5,0)
};
\addplot [only marks, error bars,y dir=both, y explicit] coordinates {
(0, 330.0) += (0, 31.0) -= (0, 24.0)
(1, 16.35) += (1, 6.6499999999999995) -= (1, 3.35)
(2, 11.350000000000001) += (2, 1.65) -= (2, 2.35)
(3, 15.65) += (3, 7.35) -= (3, 4.6499999999999995)
(4, 28.35) += (4, 5.6499999999999995) -= (4, 5.35)
(5, 7.0) += (5, 1.0) -= (5, 2.0)
(6, 6.6499999999999995) += (6, 4.35) -= (6, 2.65)
(7, 6.35) += (7, 5.6499999999999995) -= (7, 6.35)
(8, 19.349999999999998) += (8, 5.6499999999999995) -= (8, 4.35)
(9, 7.0) += (9, 1.0) -= (9, 1.0)
(10, 11.65) += (10, 1.35) -= (10, 1.65)
(11, 7.6499999999999995) += (11, 1.35) -= (11, 2.65)
(12, 12.65) += (12, 3.35) -= (12, 3.65)
(13, 9.0) += (13, 1.0) -= (13, 1.0)
(14, 6.6499999999999995) += (14, 1.35) -= (14, 0.65)
(15, 4.35) += (15, 0.65) -= (15, 1.35)
};
\addplot [dashed, mark=none] coordinates {
   (-0.5,0)
   (-0.5,500.0-38.72983346207417)
   (0.5,500.0-38.72983346207417)
   (0.5,0.0-0.0)
   (1.5,0.0-0.0)
   (1.5,0.0-0.0)
   (2.5,0.0-0.0)
   (2.5,0.0-0.0)
   (3.5,0.0-0.0)
   (3.5,0.0-0.0)
   (4.5,0.0-0.0)
   (4.5,0.0-0.0)
   (5.5,0.0-0.0)
   (5.5,0.0-0.0)
   (6.5,0.0-0.0)
   (6.5,0.0-0.0)
   (7.5,0.0-0.0)
   (7.5,0.0-0.0)
   (8.5,0.0-0.0)
   (8.5,0.0-0.0)
   (9.5,0.0-0.0)
   (9.5,0.0-0.0)
   (10.5,0.0-0.0)
   (10.5,0.0-0.0)
   (11.5,0.0-0.0)
   (11.5,0.0-0.0)
   (12.5,0.0-0.0)
   (12.5,0.0-0.0)
   (13.5,0.0-0.0)
   (13.5,0.0-0.0)
   (14.5,0.0-0.0)
   (14.5,0.0-0.0)
   (15.5,0.0-0.0)
   (15.5,0)
};
\addlegendentry{4-Wojter-2ac-pu}
\addlegendentry {3 RMS Error, 3$\times$500 shots}
\addlegendentry{Honeywell {\bf 66.0\%}} 
\end{axis}
\end{tikzpicture}
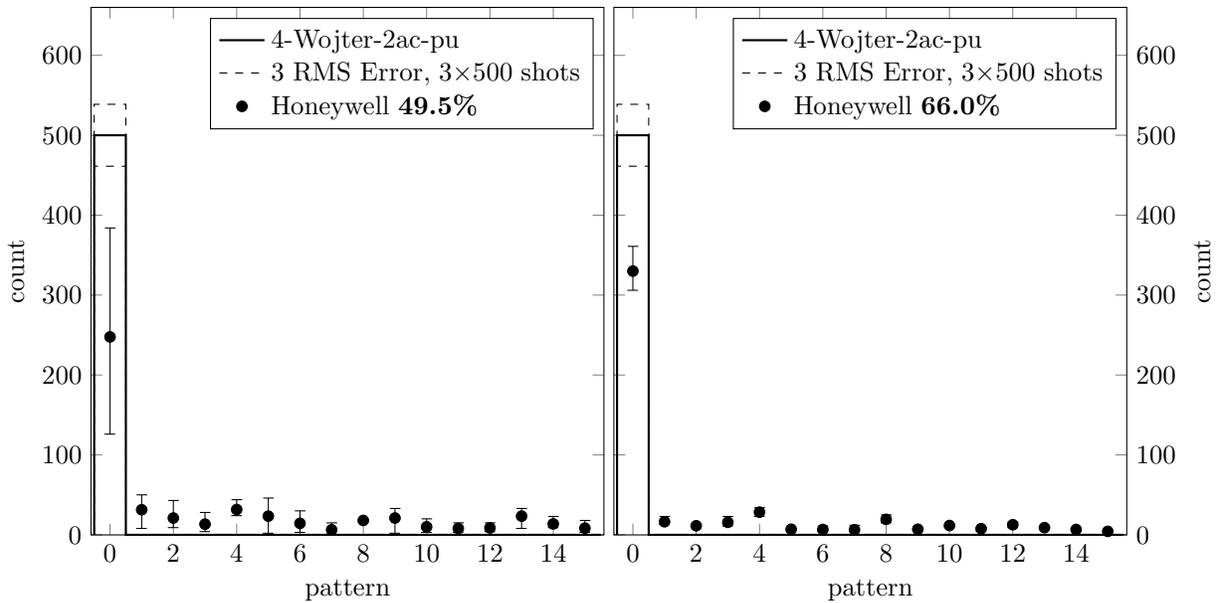
\caption{4-qubit Wojter with 2 ancilla qubits and partial uncompute: session 1 (left), session 2 (right).}
\end{ffigure}

\begin{ffigure}
\begin{tikzpicture}
\begin{axis}[
    height=8.6cm,  
    width=8.4cm,
    xmin=-0.6,xmax=15.6,ymin=-0.1,ymax=560,
    legend cell align=left,
    ylabel ={count}, ylabel near ticks, yticklabel pos=left,
    xlabel={pattern},
    xlabel near ticks, xticklabel pos=bottom,
]
\addplot [thick, mark=none] coordinates {
   (-0.5,0)
   (-0.5,406.25)
   (0.5,406.25)
   (0.5,6.25)
   (1.5,6.25)
   (1.5,6.25)
   (2.5,6.25)
   (2.5,6.25)
   (3.5,6.25)
   (3.5,6.25)
   (4.5,6.25)
   (4.5,6.25)
   (5.5,6.25)
   (5.5,6.25)
   (6.5,6.25)
   (6.5,6.25)
   (7.5,6.25)
   (15.5,0)
};
\addplot [dashed, mark=none] coordinates { 
   (-0.5,0)
   (-0.5,406.25+27.04163456597992)
   (0.5,406.25+27.04163456597992)
   (0.5,6.25+3.3541019662496847)
   (1.5,6.25+3.3541019662496847)
   (1.5,6.25+3.3541019662496847)
   (2.5,6.25+3.3541019662496847)
   (2.5,6.25+3.3541019662496847)
   (3.5,6.25+3.3541019662496847)
   (3.5,6.25+3.3541019662496847)
   (4.5,6.25+3.3541019662496847)
   (4.5,6.25+3.3541019662496847)
   (5.5,6.25+3.3541019662496847)
   (5.5,6.25+3.3541019662496847)
   (6.5,6.25+3.3541019662496847)
   (6.5,6.25+3.3541019662496847)
   (7.5,6.25+3.3541019662496847)
   (15.5,0)
};
\addplot [only marks, error bars,y dir=both, y explicit] coordinates {
(0, 195.39999999999998) += (0, 36.6) -= (0, 47.4)
(1, 13.2) += (1, 7.8) -= (1, 13.2)
(2, 15.2) += (2, 4.8) -= (2, 7.2)
(3, 7.0) += (3, 3.0) -= (3, 2.0)
(4, 30.200000000000003) += (4, 36.8) -= (4, 21.2)
(5, 11.4) += (5, 9.6) -= (5, 7.4)
(6, 19.0) += (6, 21.0) -= (6, 13.0)
(7, 15.8) += (7, 17.2) -= (7, 11.799999999999999)
(8, 42.2) += (8, 12.8) -= (8, 18.2)
(9, 30.4) += (9, 12.6) -= (9, 11.4)
(10, 32.4) += (10, 15.6) -= (10, 11.4)
(11, 30.4) += (11, 11.6) -= (11, 9.4)
(12, 16.0) += (12, 13.0) -= (12, 11.0)
(13, 16.6) += (13, 16.400000000000002) -= (13, 8.6)
(14, 13.0) += (14, 22.0) -= (14, 10.0)
(15, 11.799999999999999) += (15, 14.200000000000001) -= (15, 7.8)
};
\addplot [dashed, mark=none] coordinates {
   (-0.5,0)
   (-0.5,406.25-27.04163456597992)
   (0.5,406.25-27.04163456597992)
   (0.5,6.25-3.3541019662496847)
   (1.5,6.25-3.3541019662496847)
   (1.5,6.25-3.3541019662496847)
   (2.5,6.25-3.3541019662496847)
   (2.5,6.25-3.3541019662496847)
   (3.5,6.25-3.3541019662496847)
   (3.5,6.25-3.3541019662496847)
   (4.5,6.25-3.3541019662496847)
   (4.5,6.25-3.3541019662496847)
   (5.5,6.25-3.3541019662496847)
   (5.5,6.25-3.3541019662496847)
   (6.5,6.25-3.3541019662496847)
   (6.5,6.25-3.3541019662496847)
   (7.5,6.25-3.3541019662496847)
   (15.5,0)
};
\addlegendentry{4-Wielomianer-1ac-1am-pu}
\addlegendentry {3 RMS Error, 5$\times$500 shots}
\addlegendentry{Honeywell {\bf 39.1\%}} 
\end{axis}
\end{tikzpicture}
\begin{tikzpicture}
\begin{axis}[
    height=8.6cm,  
    width=8.4cm,
    xmin=-0.6,xmax=15.6,ymin=-0.1,ymax=560,
    legend cell align=left,
    ylabel ={count}, ylabel near ticks, yticklabel pos=right,
    xlabel={pattern},
    xlabel near ticks, xticklabel pos=bottom,
]
\addplot [thick, mark=none] coordinates {
   (-0.5,0)
   (-0.5,383.0)
   (0.5,383.0)
   (0.5,31.0)
   (1.5,31.0)
   (1.5,31.0)
   (2.5,31.0)
   (2.5,31.0)
   (3.5,31.0)
   (3.5,8.0)
   (4.5,8.0)
   (4.5,0.0)
   (5.5,0.0)
   (5.5,0.0)
   (6.5,0.0)
   (6.5,0.0)
   (7.5,0.0)
   (7.5,8.0)
   (8.5,8.0)
   (8.5,0.0)
   (9.5,0.0)
   (9.5,0.0)
   (10.5,0.0)
   (10.5,0.0)
   (11.5,0.0)
   (11.5,8.0)
   (12.5,8.0)
   (12.5,0.0)
   (13.5,0.0)
   (13.5,0.0)
   (14.5,0.0)
   (14.5,0.0)
   (15.5,0.0)
   (15.5,0)
};
\addplot [dashed, mark=none] coordinates { 
   (-0.5,0)
   (-0.5,383.0+33.896902513356586)
   (0.5,383.0+33.896902513356586)
   (0.5,31.0+9.643650760992955)
   (1.5,31.0+9.643650760992955)
   (1.5,31.0+9.643650760992955)
   (2.5,31.0+9.643650760992955)
   (2.5,31.0+9.643650760992955)
   (3.5,31.0+9.643650760992955)
   (3.5,8.0+4.898979485566357)
   (4.5,8.0+4.898979485566357)
   (4.5,0.0+0.0)
   (5.5,0.0+0.0)
   (5.5,0.0+0.0)
   (6.5,0.0+0.0)
   (6.5,0.0+0.0)
   (7.5,0.0+0.0)
   (7.5,8.0+4.898979485566357)
   (8.5,8.0+4.898979485566357)
   (8.5,0.0+0.0)
   (9.5,0.0+0.0)
   (9.5,0.0+0.0)
   (10.5,0.0+0.0)
   (10.5,0.0+0.0)
   (11.5,0.0+0.0)
   (11.5,8.0+4.898979485566357)
   (12.5,8.0+4.898979485566357)
   (12.5,0.0+0.0)
   (13.5,0.0+0.0)
   (13.5,0.0+0.0)
   (14.5,0.0+0.0)
   (14.5,0.0+0.0)
   (15.5,0.0+0.0)
   (15.5,0)
};
\addplot [only marks, error bars,y dir=both, y explicit] coordinates {
(0, 217.35) += (0, 63.65) -= (0, 53.35)
(1, 23.349999999999998) += (1, 5.6499999999999995) -= (1, 9.350000000000001)
(2, 28.0) += (2, 13.0) -= (2, 18.0)
(3, 21.65) += (3, 8.35) -= (3, 7.6499999999999995)
(4, 34.65) += (4, 17.35) -= (4, 15.65)
(5, 17.0) += (5, 7.0) -= (5, 10.0)
(6, 12.0) += (6, 7.0) -= (6, 6.0)
(7, 11.65) += (7, 4.35) -= (7, 3.65)
(8, 26.349999999999998) += (8, 25.65) -= (8, 15.350000000000001)
(9, 21.65) += (9, 16.35) -= (9, 17.65)
(10, 12.0) += (10, 12.0) -= (10, 7.0)
(11, 12.65) += (11, 4.35) -= (11, 2.65)
(12, 15.350000000000001) += (12, 2.65) -= (12, 1.35)
(13, 18.35) += (13, 12.65) -= (13, 14.35)
(14, 14.35) += (14, 8.65) -= (14, 4.35)
(15, 13.65) += (15, 4.35) -= (15, 4.6499999999999995)
};
\addplot [dashed, mark=none] coordinates {
   (-0.5,0)
   (-0.5,383.0-33.896902513356586)
   (0.5,383.0-33.896902513356586)
   (0.5,31.0-9.643650760992955)
   (1.5,31.0-9.643650760992955)
   (1.5,31.0-9.643650760992955)
   (2.5,31.0-9.643650760992955)
   (2.5,31.0-9.643650760992955)
   (3.5,31.0-9.643650760992955)
   (3.5,8.0-4.898979485566357)
   (4.5,8.0-4.898979485566357)
   (4.5,0.0-0.0)
   (5.5,0.0-0.0)
   (5.5,0.0-0.0)
   (6.5,0.0-0.0)
   (6.5,0.0-0.0)
   (7.5,0.0-0.0)
   (7.5,8.0-4.898979485566357)
   (8.5,8.0-4.898979485566357)
   (8.5,0.0-0.0)
   (9.5,0.0-0.0)
   (9.5,0.0-0.0)
   (10.5,0.0-0.0)
   (10.5,0.0-0.0)
   (11.5,0.0-0.0)
   (11.5,8.0-4.898979485566357)
   (12.5,8.0-4.898979485566357)
   (12.5,0.0-0.0)
   (13.5,0.0-0.0)
   (13.5,0.0-0.0)
   (14.5,0.0-0.0)
   (14.5,0.0-0.0)
   (15.5,0.0-0.0)
   (15.5,0)
};
\addlegendentry{4-Drzewker-2ac-pu}
\addlegendentry {3 RMS Error, 3$\times$500 shots}
\addlegendentry{Honeywell {\bf 43.5\%}} 
\end{axis}
\end{tikzpicture}
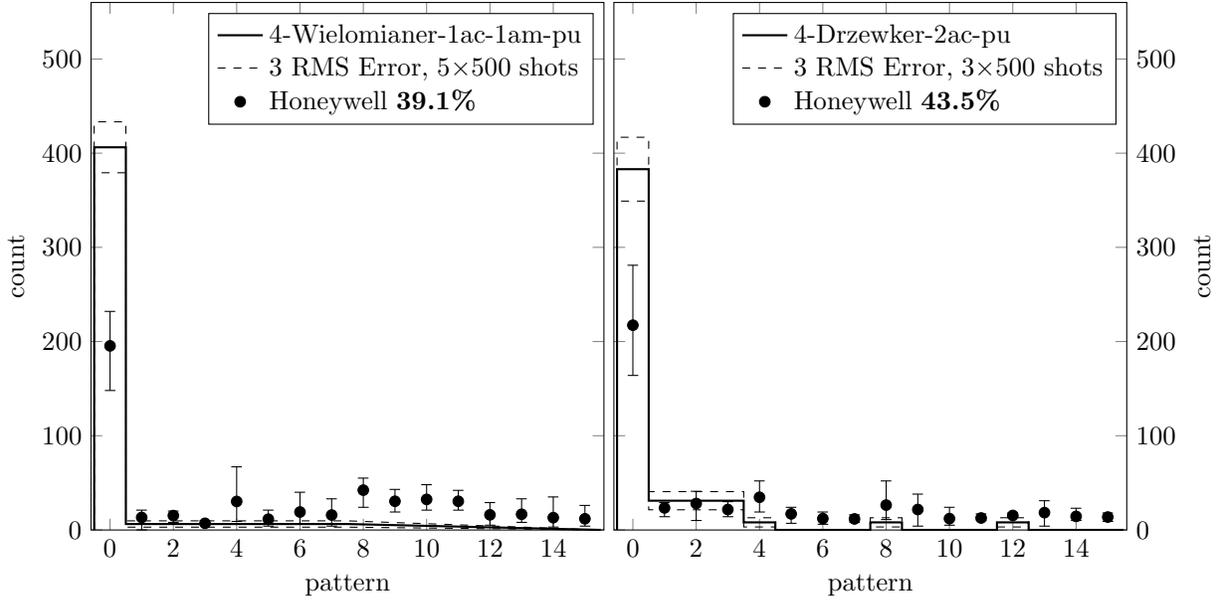
\caption{4-qubit Wielomianer with mid-circuit measurement (left) and Drzewker with 2 ancillae (right).}
\end{ffigure}

\subsection{5-qubit search}

\begin{ffigure}
\begin{tikzpicture}
\begin{axis}[
    height=8.6cm,  
    width=8.4cm,
    xmin=-0.6,xmax=31.6,ymin=-0.1,ymax=260,
    legend cell align=left,
    ylabel ={count}, ylabel near ticks, yticklabel pos=left,
    xlabel={pattern},
    xlabel near ticks, xticklabel pos=bottom,
]
\addplot [thick, mark=none] coordinates {
   (-0.5,0)
   (-0.5,129.0)
   (0.5,129.0)
   (0.5,12.0)
   (1.5,12.0)
   (1.5,12.0)
   (2.5,12.0)
   (2.5,12.0)
   (3.5,12.0)
   (3.5,12.0)
   (4.5,12.0)
   (4.5,12.0)
   (5.5,12.0)
   (5.5,12.0)
   (6.5,12.0)
   (6.5,12.0)
   (7.5,12.0)
   (7.5,12.0)
   (8.5,12.0)
   (8.5,12.0)
   (9.5,12.0)
   (9.5,12.0)
   (10.5,12.0)
   (10.5,12.0)
   (11.5,12.0)
   (11.5,12.0)
   (12.5,12.0)
   (12.5,12.0)
   (13.5,12.0)
   (13.5,12.0)
   (14.5,12.0)
   (14.5,12.0)
   (15.5,12.0)
   (15.5,12.0)
   (16.5,12.0)
   (16.5,12.0)
   (17.5,12.0)
   (17.5,12.0)
   (18.5,12.0)
   (18.5,12.0)
   (19.5,12.0)
   (19.5,12.0)
   (20.5,12.0)
   (20.5,12.0)
   (21.5,12.0)
   (21.5,12.0)
   (22.5,12.0)
   (22.5,12.0)
   (23.5,12.0)
   (23.5,12.0)
   (24.5,12.0)
   (24.5,12.0)
   (25.5,12.0)
   (25.5,12.0)
   (26.5,12.0)
   (26.5,12.0)
   (27.5,12.0)
   (27.5,12.0)
   (28.5,12.0)
   (28.5,12.0)
   (29.5,12.0)
   (29.5,12.0)
   (30.5,12.0)
   (30.5,12.0)
   (31.5,12.0)
   (31.5,0)
};
\addplot [dashed, mark=none] coordinates { 
   (-0.5,0)
   (-0.5,129.0+19.672315572906005)
   (0.5,129.0+19.672315572906005)
   (0.5,12.0+6.0)
   (1.5,12.0+6.0)
   (1.5,12.0+6.0)
   (2.5,12.0+6.0)
   (2.5,12.0+6.0)
   (3.5,12.0+6.0)
   (3.5,12.0+6.0)
   (4.5,12.0+6.0)
   (4.5,12.0+6.0)
   (5.5,12.0+6.0)
   (5.5,12.0+6.0)
   (6.5,12.0+6.0)
   (6.5,12.0+6.0)
   (7.5,12.0+6.0)
   (7.5,12.0+6.0)
   (8.5,12.0+6.0)
   (8.5,12.0+6.0)
   (9.5,12.0+6.0)
   (9.5,12.0+6.0)
   (10.5,12.0+6.0)
   (10.5,12.0+6.0)
   (11.5,12.0+6.0)
   (11.5,12.0+6.0)
   (12.5,12.0+6.0)
   (12.5,12.0+6.0)
   (13.5,12.0+6.0)
   (13.5,12.0+6.0)
   (14.5,12.0+6.0)
   (14.5,12.0+6.0)
   (15.5,12.0+6.0)
   (15.5,12.0+6.0)
   (16.5,12.0+6.0)
   (16.5,12.0+6.0)
   (17.5,12.0+6.0)
   (17.5,12.0+6.0)
   (18.5,12.0+6.0)
   (18.5,12.0+6.0)
   (19.5,12.0+6.0)
   (19.5,12.0+6.0)
   (20.5,12.0+6.0)
   (20.5,12.0+6.0)
   (21.5,12.0+6.0)
   (21.5,12.0+6.0)
   (22.5,12.0+6.0)
   (22.5,12.0+6.0)
   (23.5,12.0+6.0)
   (23.5,12.0+6.0)
   (24.5,12.0+6.0)
   (24.5,12.0+6.0)
   (25.5,12.0+6.0)
   (25.5,12.0+6.0)
   (26.5,12.0+6.0)
   (26.5,12.0+6.0)
   (27.5,12.0+6.0)
   (27.5,12.0+6.0)
   (28.5,12.0+6.0)
   (28.5,12.0+6.0)
   (29.5,12.0+6.0)
   (29.5,12.0+6.0)
   (30.5,12.0+6.0)
   (30.5,12.0+6.0)
   (31.5,12.0+6.0)
   (31.5,0)
};
\addplot [only marks, error bars,y dir=both, y explicit] coordinates {
(0, 93.35000000000001) += (0, 11.65) -= (0, 9.350000000000001)
(1, 11.350000000000001) += (1, 4.6499999999999995) -= (1, 4.35)
(2, 16.35) += (2, 1.65) -= (2, 3.35)
(3, 7.35) += (3, 1.65) -= (3, 2.35)
(4, 10.35) += (4, 2.65) -= (4, 2.35)
(5, 5.6499999999999995) += (5, 1.35) -= (5, 2.65)
(6, 5.0) += (6, 1.0) -= (6, 1.0)
(7, 12.0) += (7, 2.0) -= (7, 3.0)
(8, 12.35) += (8, 1.65) -= (8, 2.35)
(9, 8.35) += (9, 3.65) -= (9, 3.35)
(10, 7.0) += (10, 2.0) -= (10, 4.0)
(11, 10.35) += (11, 3.65) -= (11, 4.35)
(12, 6.35) += (12, 1.65) -= (12, 1.35)
(13, 9.350000000000001) += (13, 3.65) -= (13, 3.35)
(14, 11.65) += (14, 0.35) -= (14, 0.65)
(15, 23.349999999999998) += (15, 2.65) -= (15, 2.35)
(16, 12.0) += (16, 5.0) -= (16, 4.0)
(17, 8.65) += (17, 5.35) -= (17, 3.65)
(18, 7.35) += (18, 3.65) -= (18, 4.35)
(19, 12.65) += (19, 3.35) -= (19, 5.6499999999999995)
(20, 9.65) += (20, 2.35) -= (20, 1.65)
(21, 13.65) += (21, 8.35) -= (21, 5.6499999999999995)
(22, 17.35) += (22, 0.65) -= (22, 1.35)
(23, 18.65) += (23, 5.35) -= (23, 2.65)
(24, 8.0) += (24, 2.0) -= (24, 3.0)
(25, 9.65) += (25, 2.35) -= (25, 2.65)
(26, 17.65) += (26, 9.350000000000001) -= (26, 6.6499999999999995)
(27, 20.0) += (27, 12.0) -= (27, 7.0)
(28, 18.35) += (28, 3.65) -= (28, 3.35)
(29, 20.0) += (29, 6.0) -= (29, 9.0)
(30, 24.0) += (30, 5.0) -= (30, 6.0)
(31, 32.349999999999994) += (31, 1.65) -= (31, 1.35)
};
\addplot [dashed, mark=none] coordinates {
   (-0.5,0)
   (-0.5,129.0-19.672315572906005)
   (0.5,129.0-19.672315572906005)
   (0.5,12.0-6.0)
   (1.5,12.0-6.0)
   (1.5,12.0-6.0)
   (2.5,12.0-6.0)
   (2.5,12.0-6.0)
   (3.5,12.0-6.0)
   (3.5,12.0-6.0)
   (4.5,12.0-6.0)
   (4.5,12.0-6.0)
   (5.5,12.0-6.0)
   (5.5,12.0-6.0)
   (6.5,12.0-6.0)
   (6.5,12.0-6.0)
   (7.5,12.0-6.0)
   (7.5,12.0-6.0)
   (8.5,12.0-6.0)
   (8.5,12.0-6.0)
   (9.5,12.0-6.0)
   (9.5,12.0-6.0)
   (10.5,12.0-6.0)
   (10.5,12.0-6.0)
   (11.5,12.0-6.0)
   (11.5,12.0-6.0)
   (12.5,12.0-6.0)
   (12.5,12.0-6.0)
   (13.5,12.0-6.0)
   (13.5,12.0-6.0)
   (14.5,12.0-6.0)
   (14.5,12.0-6.0)
   (15.5,12.0-6.0)
   (15.5,12.0-6.0)
   (16.5,12.0-6.0)
   (16.5,12.0-6.0)
   (17.5,12.0-6.0)
   (17.5,12.0-6.0)
   (18.5,12.0-6.0)
   (18.5,12.0-6.0)
   (19.5,12.0-6.0)
   (19.5,12.0-6.0)
   (20.5,12.0-6.0)
   (20.5,12.0-6.0)
   (21.5,12.0-6.0)
   (21.5,12.0-6.0)
   (22.5,12.0-6.0)
   (22.5,12.0-6.0)
   (23.5,12.0-6.0)
   (23.5,12.0-6.0)
   (24.5,12.0-6.0)
   (24.5,12.0-6.0)
   (25.5,12.0-6.0)
   (25.5,12.0-6.0)
   (26.5,12.0-6.0)
   (26.5,12.0-6.0)
   (27.5,12.0-6.0)
   (27.5,12.0-6.0)
   (28.5,12.0-6.0)
   (28.5,12.0-6.0)
   (29.5,12.0-6.0)
   (29.5,12.0-6.0)
   (30.5,12.0-6.0)
   (30.5,12.0-6.0)
   (31.5,12.0-6.0)
   (31.5,0)
};
\addlegendentry{5-Grover-1ac-a}
\addlegendentry {3 RMS Error, 3$\times$500 shots}
\addlegendentry{Honeywell {\bf 18.7\%}} 
\end{axis}
\end{tikzpicture}
\begin{tikzpicture}
\begin{axis}[
    height=8.6cm,  
    width=8.4cm,
    xmin=-0.6,xmax=31.6,ymin=-0.1,ymax=260,
    legend cell align=left,
    ylabel ={count}, ylabel near ticks, yticklabel pos=right,
    xlabel={pattern},
    xlabel near ticks, xticklabel pos=bottom,
]
\addplot [thick, mark=none] coordinates {
   (-0.5,0)
   (-0.5,129.0)
   (0.5,129.0)
   (0.5,12.0)
   (1.5,12.0)
   (1.5,12.0)
   (2.5,12.0)
   (2.5,12.0)
   (3.5,12.0)
   (3.5,12.0)
   (4.5,12.0)
   (4.5,12.0)
   (5.5,12.0)
   (5.5,12.0)
   (6.5,12.0)
   (6.5,12.0)
   (7.5,12.0)
   (7.5,12.0)
   (8.5,12.0)
   (8.5,12.0)
   (9.5,12.0)
   (9.5,12.0)
   (10.5,12.0)
   (10.5,12.0)
   (11.5,12.0)
   (11.5,12.0)
   (12.5,12.0)
   (12.5,12.0)
   (13.5,12.0)
   (13.5,12.0)
   (14.5,12.0)
   (14.5,12.0)
   (15.5,12.0)
   (15.5,12.0)
   (16.5,12.0)
   (16.5,12.0)
   (17.5,12.0)
   (17.5,12.0)
   (18.5,12.0)
   (18.5,12.0)
   (19.5,12.0)
   (19.5,12.0)
   (20.5,12.0)
   (20.5,12.0)
   (21.5,12.0)
   (21.5,12.0)
   (22.5,12.0)
   (22.5,12.0)
   (23.5,12.0)
   (23.5,12.0)
   (24.5,12.0)
   (24.5,12.0)
   (25.5,12.0)
   (25.5,12.0)
   (26.5,12.0)
   (26.5,12.0)
   (27.5,12.0)
   (27.5,12.0)
   (28.5,12.0)
   (28.5,12.0)
   (29.5,12.0)
   (29.5,12.0)
   (30.5,12.0)
   (30.5,12.0)
   (31.5,12.0)
   (31.5,0)
};
\addplot [dashed, mark=none] coordinates { 
   (-0.5,0)
   (-0.5,129.0+19.672315572906005)
   (0.5,129.0+19.672315572906005)
   (0.5,12.0+6.0)
   (1.5,12.0+6.0)
   (1.5,12.0+6.0)
   (2.5,12.0+6.0)
   (2.5,12.0+6.0)
   (3.5,12.0+6.0)
   (3.5,12.0+6.0)
   (4.5,12.0+6.0)
   (4.5,12.0+6.0)
   (5.5,12.0+6.0)
   (5.5,12.0+6.0)
   (6.5,12.0+6.0)
   (6.5,12.0+6.0)
   (7.5,12.0+6.0)
   (7.5,12.0+6.0)
   (8.5,12.0+6.0)
   (8.5,12.0+6.0)
   (9.5,12.0+6.0)
   (9.5,12.0+6.0)
   (10.5,12.0+6.0)
   (10.5,12.0+6.0)
   (11.5,12.0+6.0)
   (11.5,12.0+6.0)
   (12.5,12.0+6.0)
   (12.5,12.0+6.0)
   (13.5,12.0+6.0)
   (13.5,12.0+6.0)
   (14.5,12.0+6.0)
   (14.5,12.0+6.0)
   (15.5,12.0+6.0)
   (15.5,12.0+6.0)
   (16.5,12.0+6.0)
   (16.5,12.0+6.0)
   (17.5,12.0+6.0)
   (17.5,12.0+6.0)
   (18.5,12.0+6.0)
   (18.5,12.0+6.0)
   (19.5,12.0+6.0)
   (19.5,12.0+6.0)
   (20.5,12.0+6.0)
   (20.5,12.0+6.0)
   (21.5,12.0+6.0)
   (21.5,12.0+6.0)
   (22.5,12.0+6.0)
   (22.5,12.0+6.0)
   (23.5,12.0+6.0)
   (23.5,12.0+6.0)
   (24.5,12.0+6.0)
   (24.5,12.0+6.0)
   (25.5,12.0+6.0)
   (25.5,12.0+6.0)
   (26.5,12.0+6.0)
   (26.5,12.0+6.0)
   (27.5,12.0+6.0)
   (27.5,12.0+6.0)
   (28.5,12.0+6.0)
   (28.5,12.0+6.0)
   (29.5,12.0+6.0)
   (29.5,12.0+6.0)
   (30.5,12.0+6.0)
   (30.5,12.0+6.0)
   (31.5,12.0+6.0)
   (31.5,0)
};
\addplot [only marks, error bars,y dir=both, y explicit] coordinates {
(0, 79.65) += (0, 18.35) -= (0, 11.65)
(1, 12.0) += (1, 3.0) -= (1, 2.0)
(2, 19.650000000000002) += (2, 4.35) -= (2, 2.65)
(3, 12.65) += (3, 4.35) -= (3, 5.6499999999999995)
(4, 9.65) += (4, 6.35) -= (4, 5.6499999999999995)
(5, 7.0) += (5, 2.0) -= (5, 2.0)
(6, 6.6499999999999995) += (6, 5.35) -= (6, 3.65)
(7, 16.650000000000002) += (7, 0.35) -= (7, 0.65)
(8, 15.0) += (8, 3.0) -= (8, 2.0)
(9, 6.6499999999999995) += (9, 3.35) -= (9, 3.65)
(10, 6.0) += (10, 4.0) -= (10, 3.0)
(11, 15.65) += (11, 3.35) -= (11, 5.6499999999999995)
(12, 8.35) += (12, 0.65) -= (12, 1.35)
(13, 14.35) += (13, 3.65) -= (13, 4.35)
(14, 11.0) += (14, 2.0) -= (14, 1.0)
(15, 25.0) += (15, 4.0) -= (15, 5.0)
(16, 11.350000000000001) += (16, 6.6499999999999995) -= (16, 5.35)
(17, 10.65) += (17, 2.35) -= (17, 1.65)
(18, 10.35) += (18, 5.6499999999999995) -= (18, 3.35)
(19, 19.0) += (19, 4.0) -= (19, 2.0)
(20, 10.65) += (20, 5.35) -= (20, 2.65)
(21, 11.0) += (21, 4.0) -= (21, 5.0)
(22, 10.65) += (22, 2.35) -= (22, 2.65)
(23, 20.35) += (23, 12.65) -= (23, 9.350000000000001)
(24, 10.0) += (24, 2.0) -= (24, 3.0)
(25, 13.65) += (25, 1.35) -= (25, 1.65)
(26, 13.65) += (26, 3.35) -= (26, 4.6499999999999995)
(27, 21.0) += (27, 10.0) -= (27, 9.0)
(28, 12.35) += (28, 4.6499999999999995) -= (28, 3.35)
(29, 19.349999999999998) += (29, 6.6499999999999995) -= (29, 7.35)
(30, 17.0) += (30, 3.0) -= (30, 4.0)
(31, 23.0) += (31, 1.0) -= (31, 2.0)
};
\addplot [dashed, mark=none] coordinates {
   (-0.5,0)
   (-0.5,129.0-19.672315572906005)
   (0.5,129.0-19.672315572906005)
   (0.5,12.0-6.0)
   (1.5,12.0-6.0)
   (1.5,12.0-6.0)
   (2.5,12.0-6.0)
   (2.5,12.0-6.0)
   (3.5,12.0-6.0)
   (3.5,12.0-6.0)
   (4.5,12.0-6.0)
   (4.5,12.0-6.0)
   (5.5,12.0-6.0)
   (5.5,12.0-6.0)
   (6.5,12.0-6.0)
   (6.5,12.0-6.0)
   (7.5,12.0-6.0)
   (7.5,12.0-6.0)
   (8.5,12.0-6.0)
   (8.5,12.0-6.0)
   (9.5,12.0-6.0)
   (9.5,12.0-6.0)
   (10.5,12.0-6.0)
   (10.5,12.0-6.0)
   (11.5,12.0-6.0)
   (11.5,12.0-6.0)
   (12.5,12.0-6.0)
   (12.5,12.0-6.0)
   (13.5,12.0-6.0)
   (13.5,12.0-6.0)
   (14.5,12.0-6.0)
   (14.5,12.0-6.0)
   (15.5,12.0-6.0)
   (15.5,12.0-6.0)
   (16.5,12.0-6.0)
   (16.5,12.0-6.0)
   (17.5,12.0-6.0)
   (17.5,12.0-6.0)
   (18.5,12.0-6.0)
   (18.5,12.0-6.0)
   (19.5,12.0-6.0)
   (19.5,12.0-6.0)
   (20.5,12.0-6.0)
   (20.5,12.0-6.0)
   (21.5,12.0-6.0)
   (21.5,12.0-6.0)
   (22.5,12.0-6.0)
   (22.5,12.0-6.0)
   (23.5,12.0-6.0)
   (23.5,12.0-6.0)
   (24.5,12.0-6.0)
   (24.5,12.0-6.0)
   (25.5,12.0-6.0)
   (25.5,12.0-6.0)
   (26.5,12.0-6.0)
   (26.5,12.0-6.0)
   (27.5,12.0-6.0)
   (27.5,12.0-6.0)
   (28.5,12.0-6.0)
   (28.5,12.0-6.0)
   (29.5,12.0-6.0)
   (29.5,12.0-6.0)
   (30.5,12.0-6.0)
   (30.5,12.0-6.0)
   (31.5,12.0-6.0)
   (31.5,0)
};
\addlegendentry{5-Grover-1ac-a}
\addlegendentry {3 RMS Error, 3$\times$500 shots}
\addlegendentry{Honeywell {\bf 15.9\%}} 
\end{axis}
\end{tikzpicture}
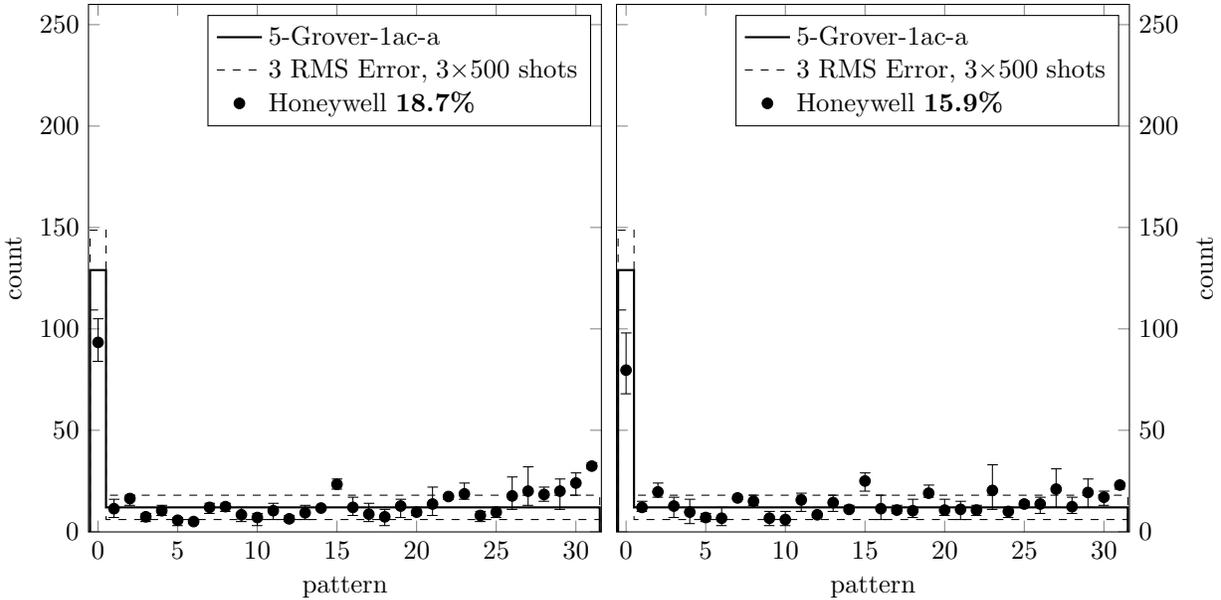
\caption{5-qubit 36CX Grover with 1 ancilla: session 1 (left), session 2 (right).}
    \label{fig:5Grover}
\end{ffigure}

\begin{ffigure}
\begin{tikzpicture}
\begin{axis}[
    height=8.6cm,  
    width=8.4cm,
    xmin=-0.6,xmax=31.6,ymin=-0.1,ymax=460,
    legend cell align=left,
    ylabel ={count}, ylabel near ticks, yticklabel pos=left,
    xlabel={pattern},
    xlabel near ticks, xticklabel pos=bottom,
]
\addplot [thick, mark=none] coordinates {
   (-0.5,0)
   (-0.5,282.0)
   (0.5,282.0)
   (0.5,9.0)
   (1.5,9.0)
   (1.5,9.0)
   (2.5,9.0)
   (2.5,9.0)
   (3.5,9.0)
   (3.5,15.5)
   (4.5,15.5)
   (4.5,4.0)
   (5.5,4.0)
   (5.5,4.0)
   (6.5,4.0)
   (6.5,4.0)
   (7.5,4.0)
   (7.5,15.5)
   (8.5,15.5)
   (8.5,4.0)
   (9.5,4.0)
   (9.5,4.0)
   (10.5,4.0)
   (10.5,4.0)
   (11.5,4.0)
   (11.5,15.5)
   (12.5,15.5)
   (12.5,4.0)
   (13.5,4.0)
   (13.5,4.0)
   (14.5,4.0)
   (14.5,4.0)
   (15.5,4.0)
   (15.5,15.5)
   (16.5,15.5)
   (16.5,4.0)
   (17.5,4.0)
   (17.5,4.0)
   (18.5,4.0)
   (18.5,4.0)
   (19.5,4.0)
   (19.5,15.5)
   (20.5,15.5)
   (20.5,4.0)
   (21.5,4.0)
   (21.5,4.0)
   (22.5,4.0)
   (22.5,4.0)
   (23.5,4.0)
   (23.5,15.5)
   (24.5,15.5)
   (24.5,4.0)
   (25.5,4.0)
   (25.5,4.0)
   (26.5,4.0)
   (26.5,4.0)
   (27.5,4.0)
   (27.5,15.5)
   (28.5,15.5)
   (28.5,4.0)
   (29.5,4.0)
   (29.5,4.0)
   (30.5,4.0)
   (30.5,4.0)
   (31.5,4.0)
   (31.5,0)
};
\addplot [dashed, mark=none] coordinates { 
   (-0.5,0)
   (-0.5,282.0+29.08607914449797)
   (0.5,282.0+29.08607914449797)
   (0.5,9.0+5.196152422706632)
   (1.5,9.0+5.196152422706632)
   (1.5,9.0+5.196152422706632)
   (2.5,9.0+5.196152422706632)
   (2.5,9.0+5.196152422706632)
   (3.5,9.0+5.196152422706632)
   (3.5,15.5+6.819090848492927)
   (4.5,15.5+6.819090848492927)
   (4.5,4.0+3.464101615137755)
   (5.5,4.0+3.464101615137755)
   (5.5,4.0+3.464101615137755)
   (6.5,4.0+3.464101615137755)
   (6.5,4.0+3.464101615137755)
   (7.5,4.0+3.464101615137755)
   (7.5,15.5+6.819090848492927)
   (8.5,15.5+6.819090848492927)
   (8.5,4.0+3.464101615137755)
   (9.5,4.0+3.464101615137755)
   (9.5,4.0+3.464101615137755)
   (10.5,4.0+3.464101615137755)
   (10.5,4.0+3.464101615137755)
   (11.5,4.0+3.464101615137755)
   (11.5,15.5+6.819090848492927)
   (12.5,15.5+6.819090848492927)
   (12.5,4.0+3.464101615137755)
   (13.5,4.0+3.464101615137755)
   (13.5,4.0+3.464101615137755)
   (14.5,4.0+3.464101615137755)
   (14.5,4.0+3.464101615137755)
   (15.5,4.0+3.464101615137755)
   (15.5,15.5+6.819090848492927)
   (16.5,15.5+6.819090848492927)
   (16.5,4.0+3.464101615137755)
   (17.5,4.0+3.464101615137755)
   (17.5,4.0+3.464101615137755)
   (18.5,4.0+3.464101615137755)
   (18.5,4.0+3.464101615137755)
   (19.5,4.0+3.464101615137755)
   (19.5,15.5+6.819090848492927)
   (20.5,15.5+6.819090848492927)
   (20.5,4.0+3.464101615137755)
   (21.5,4.0+3.464101615137755)
   (21.5,4.0+3.464101615137755)
   (22.5,4.0+3.464101615137755)
   (22.5,4.0+3.464101615137755)
   (23.5,4.0+3.464101615137755)
   (23.5,15.5+6.819090848492927)
   (24.5,15.5+6.819090848492927)
   (24.5,4.0+3.464101615137755)
   (25.5,4.0+3.464101615137755)
   (25.5,4.0+3.464101615137755)
   (26.5,4.0+3.464101615137755)
   (26.5,4.0+3.464101615137755)
   (27.5,4.0+3.464101615137755)
   (27.5,15.5+6.819090848492927)
   (28.5,15.5+6.819090848492927)
   (28.5,4.0+3.464101615137755)
   (29.5,4.0+3.464101615137755)
   (29.5,4.0+3.464101615137755)
   (30.5,4.0+3.464101615137755)
   (30.5,4.0+3.464101615137755)
   (31.5,4.0+3.464101615137755)
   (31.5,0)
};
\addplot [only marks, error bars,y dir=both, y explicit] coordinates {
(0, 82.35000000000001) += (0, 39.65) -= (0, 30.349999999999998)
(1, 68.0) += (1, 34.0) -= (1, 40.0)
(2, 27.650000000000002) += (2, 8.35) -= (2, 8.65)
(3, 17.35) += (3, 9.65) -= (3, 8.35)
(4, 13.0) += (4, 6.0) -= (4, 9.0)
(5, 9.0) += (5, 3.0) -= (5, 5.0)
(6, 13.350000000000001) += (6, 4.6499999999999995) -= (6, 6.35)
(7, 8.0) += (7, 1.0) -= (7, 1.0)
(8, 13.350000000000001) += (8, 5.6499999999999995) -= (8, 9.350000000000001)
(9, 11.65) += (9, 8.35) -= (9, 5.6499999999999995)
(10, 11.65) += (10, 7.35) -= (10, 8.65)
(11, 7.6499999999999995) += (11, 3.35) -= (11, 5.6499999999999995)
(12, 10.35) += (12, 1.65) -= (12, 3.35)
(13, 6.0) += (13, 6.0) -= (13, 4.0)
(14, 9.350000000000001) += (14, 1.65) -= (14, 2.35)
(15, 7.35) += (15, 8.65) -= (15, 6.35)
(16, 18.35) += (16, 8.65) -= (16, 8.35)
(17, 16.35) += (17, 14.65) -= (17, 8.35)
(18, 10.0) += (18, 6.0) -= (18, 8.0)
(19, 8.35) += (19, 6.6499999999999995) -= (19, 3.35)
(20, 11.65) += (20, 4.35) -= (20, 6.6499999999999995)
(21, 8.65) += (21, 4.35) -= (21, 5.6499999999999995)
(22, 6.0) += (22, 4.0) -= (22, 5.0)
(23, 5.0) += (23, 6.0) -= (23, 5.0)
(24, 15.0) += (24, 7.0) -= (24, 8.0)
(25, 6.35) += (25, 8.65) -= (25, 5.35)
(26, 7.6499999999999995) += (26, 6.35) -= (26, 3.65)
(27, 4.6499999999999995) += (27, 6.35) -= (27, 3.65)
(28, 25.35) += (28, 28.65) -= (28, 22.349999999999998)
(29, 15.350000000000001) += (29, 11.65) -= (29, 8.35)
(30, 17.35) += (30, 25.65) -= (30, 15.350000000000001)
(31, 8.0) += (31, 5.0) -= (31, 7.0)
};
\addplot [dashed, mark=none] coordinates {
   (-0.5,0)
   (-0.5,282.0-29.08607914449797)
   (0.5,282.0-29.08607914449797)
   (0.5,9.0-5.196152422706632)
   (1.5,9.0-5.196152422706632)
   (1.5,9.0-5.196152422706632)
   (2.5,9.0-5.196152422706632)
   (2.5,9.0-5.196152422706632)
   (3.5,9.0-5.196152422706632)
   (3.5,15.5-6.819090848492927)
   (4.5,15.5-6.819090848492927)
   (4.5,4.0-3.464101615137755)
   (5.5,4.0-3.464101615137755)
   (5.5,4.0-3.464101615137755)
   (6.5,4.0-3.464101615137755)
   (6.5,4.0-3.464101615137755)
   (7.5,4.0-3.464101615137755)
   (7.5,15.5-6.819090848492927)
   (8.5,15.5-6.819090848492927)
   (8.5,4.0-3.464101615137755)
   (9.5,4.0-3.464101615137755)
   (9.5,4.0-3.464101615137755)
   (10.5,4.0-3.464101615137755)
   (10.5,4.0-3.464101615137755)
   (11.5,4.0-3.464101615137755)
   (11.5,15.5-6.819090848492927)
   (12.5,15.5-6.819090848492927)
   (12.5,4.0-3.464101615137755)
   (13.5,4.0-3.464101615137755)
   (13.5,4.0-3.464101615137755)
   (14.5,4.0-3.464101615137755)
   (14.5,4.0-3.464101615137755)
   (15.5,4.0-3.464101615137755)
   (15.5,15.5-6.819090848492927)
   (16.5,15.5-6.819090848492927)
   (16.5,4.0-3.464101615137755)
   (17.5,4.0-3.464101615137755)
   (17.5,4.0-3.464101615137755)
   (18.5,4.0-3.464101615137755)
   (18.5,4.0-3.464101615137755)
   (19.5,4.0-3.464101615137755)
   (19.5,15.5-6.819090848492927)
   (20.5,15.5-6.819090848492927)
   (20.5,4.0-3.464101615137755)
   (21.5,4.0-3.464101615137755)
   (21.5,4.0-3.464101615137755)
   (22.5,4.0-3.464101615137755)
   (22.5,4.0-3.464101615137755)
   (23.5,4.0-3.464101615137755)
   (23.5,15.5-6.819090848492927)
   (24.5,15.5-6.819090848492927)
   (24.5,4.0-3.464101615137755)
   (25.5,4.0-3.464101615137755)
   (25.5,4.0-3.464101615137755)
   (26.5,4.0-3.464101615137755)
   (26.5,4.0-3.464101615137755)
   (27.5,4.0-3.464101615137755)
   (27.5,15.5-6.819090848492927)
   (28.5,15.5-6.819090848492927)
   (28.5,4.0-3.464101615137755)
   (29.5,4.0-3.464101615137755)
   (29.5,4.0-3.464101615137755)
   (30.5,4.0-3.464101615137755)
   (30.5,4.0-3.464101615137755)
   (31.5,4.0-3.464101615137755)
   (31.5,0)
};
\addlegendentry{5-Drzewker-1ac-pu-c}
\addlegendentry {3 RMS Error, 3$\times$500 shots}
\addlegendentry{Honeywell {\bf 16.5\%}} 
\end{axis}
\end{tikzpicture}
\begin{tikzpicture}
\begin{axis}[
    height=8.6cm,  
    width=8.4cm,
    xmin=-0.6,xmax=31.6,ymin=-0.1,ymax=460,
    legend cell align=left,
    ylabel ={count}, ylabel near ticks, yticklabel pos=right,
    xlabel={pattern},
    xlabel near ticks, xticklabel pos=bottom,
]
\addplot [thick, mark=none] coordinates {
   (-0.5,0)
   (-0.5,282.0)
   (0.5,282.0)
   (0.5,9.0)
   (1.5,9.0)
   (1.5,9.0)
   (2.5,9.0)
   (2.5,9.0)
   (3.5,9.0)
   (3.5,15.5)
   (4.5,15.5)
   (4.5,4.0)
   (5.5,4.0)
   (5.5,4.0)
   (6.5,4.0)
   (6.5,4.0)
   (7.5,4.0)
   (7.5,15.5)
   (8.5,15.5)
   (8.5,4.0)
   (9.5,4.0)
   (9.5,4.0)
   (10.5,4.0)
   (10.5,4.0)
   (11.5,4.0)
   (11.5,15.5)
   (12.5,15.5)
   (12.5,4.0)
   (13.5,4.0)
   (13.5,4.0)
   (14.5,4.0)
   (14.5,4.0)
   (15.5,4.0)
   (15.5,15.5)
   (16.5,15.5)
   (16.5,4.0)
   (17.5,4.0)
   (17.5,4.0)
   (18.5,4.0)
   (18.5,4.0)
   (19.5,4.0)
   (19.5,15.5)
   (20.5,15.5)
   (20.5,4.0)
   (21.5,4.0)
   (21.5,4.0)
   (22.5,4.0)
   (22.5,4.0)
   (23.5,4.0)
   (23.5,15.5)
   (24.5,15.5)
   (24.5,4.0)
   (25.5,4.0)
   (25.5,4.0)
   (26.5,4.0)
   (26.5,4.0)
   (27.5,4.0)
   (27.5,15.5)
   (28.5,15.5)
   (28.5,4.0)
   (29.5,4.0)
   (29.5,4.0)
   (30.5,4.0)
   (30.5,4.0)
   (31.5,4.0)
   (31.5,0)
};
\addplot [dashed, mark=none] coordinates { 
   (-0.5,0)
   (-0.5,282.0+29.08607914449797)
   (0.5,282.0+29.08607914449797)
   (0.5,9.0+5.196152422706632)
   (1.5,9.0+5.196152422706632)
   (1.5,9.0+5.196152422706632)
   (2.5,9.0+5.196152422706632)
   (2.5,9.0+5.196152422706632)
   (3.5,9.0+5.196152422706632)
   (3.5,15.5+6.819090848492927)
   (4.5,15.5+6.819090848492927)
   (4.5,4.0+3.464101615137755)
   (5.5,4.0+3.464101615137755)
   (5.5,4.0+3.464101615137755)
   (6.5,4.0+3.464101615137755)
   (6.5,4.0+3.464101615137755)
   (7.5,4.0+3.464101615137755)
   (7.5,15.5+6.819090848492927)
   (8.5,15.5+6.819090848492927)
   (8.5,4.0+3.464101615137755)
   (9.5,4.0+3.464101615137755)
   (9.5,4.0+3.464101615137755)
   (10.5,4.0+3.464101615137755)
   (10.5,4.0+3.464101615137755)
   (11.5,4.0+3.464101615137755)
   (11.5,15.5+6.819090848492927)
   (12.5,15.5+6.819090848492927)
   (12.5,4.0+3.464101615137755)
   (13.5,4.0+3.464101615137755)
   (13.5,4.0+3.464101615137755)
   (14.5,4.0+3.464101615137755)
   (14.5,4.0+3.464101615137755)
   (15.5,4.0+3.464101615137755)
   (15.5,15.5+6.819090848492927)
   (16.5,15.5+6.819090848492927)
   (16.5,4.0+3.464101615137755)
   (17.5,4.0+3.464101615137755)
   (17.5,4.0+3.464101615137755)
   (18.5,4.0+3.464101615137755)
   (18.5,4.0+3.464101615137755)
   (19.5,4.0+3.464101615137755)
   (19.5,15.5+6.819090848492927)
   (20.5,15.5+6.819090848492927)
   (20.5,4.0+3.464101615137755)
   (21.5,4.0+3.464101615137755)
   (21.5,4.0+3.464101615137755)
   (22.5,4.0+3.464101615137755)
   (22.5,4.0+3.464101615137755)
   (23.5,4.0+3.464101615137755)
   (23.5,15.5+6.819090848492927)
   (24.5,15.5+6.819090848492927)
   (24.5,4.0+3.464101615137755)
   (25.5,4.0+3.464101615137755)
   (25.5,4.0+3.464101615137755)
   (26.5,4.0+3.464101615137755)
   (26.5,4.0+3.464101615137755)
   (27.5,4.0+3.464101615137755)
   (27.5,15.5+6.819090848492927)
   (28.5,15.5+6.819090848492927)
   (28.5,4.0+3.464101615137755)
   (29.5,4.0+3.464101615137755)
   (29.5,4.0+3.464101615137755)
   (30.5,4.0+3.464101615137755)
   (30.5,4.0+3.464101615137755)
   (31.5,4.0+3.464101615137755)
   (31.5,0)
};
\addplot [only marks, error bars,y dir=both, y explicit] coordinates {
(0, 128.0) += (0, 13.0) -= (0, 10.0)
(1, 21.65) += (1, 7.35) -= (1, 6.6499999999999995)
(2, 24.65) += (2, 3.35) -= (2, 2.65)
(3, 14.0) += (3, 7.0) -= (3, 4.0)
(4, 14.65) += (4, 6.35) -= (4, 3.65)
(5, 6.6499999999999995) += (5, 2.35) -= (5, 3.65)
(6, 8.35) += (6, 4.6499999999999995) -= (6, 2.35)
(7, 7.6499999999999995) += (7, 3.35) -= (7, 2.65)
(8, 19.0) += (8, 2.0) -= (8, 2.0)
(9, 9.0) += (9, 9.0) -= (9, 8.0)
(10, 7.6499999999999995) += (10, 1.35) -= (10, 1.65)
(11, 10.65) += (11, 2.35) -= (11, 1.65)
(12, 20.0) += (12, 4.0) -= (12, 3.0)
(13, 9.65) += (13, 3.35) -= (13, 3.65)
(14, 3.65) += (14, 0.35) -= (14, 0.65)
(15, 6.35) += (15, 1.65) -= (15, 1.35)
(16, 24.65) += (16, 5.35) -= (16, 4.6499999999999995)
(17, 15.350000000000001) += (17, 8.65) -= (17, 5.35)
(18, 13.0) += (18, 0.0) -= (18, 0.0)
(19, 15.0) += (19, 5.0) -= (19, 7.0)
(20, 15.65) += (20, 1.35) -= (20, 1.65)
(21, 6.6499999999999995) += (21, 1.35) -= (21, 2.65)
(22, 6.0) += (22, 3.0) -= (22, 4.0)
(23, 6.6499999999999995) += (23, 1.35) -= (23, 1.65)
(24, 16.0) += (24, 2.0) -= (24, 3.0)
(25, 5.35) += (25, 3.65) -= (25, 2.35)
(26, 7.0) += (26, 3.0) -= (26, 4.0)
(27, 9.0) += (27, 1.0) -= (27, 2.0)
(28, 20.650000000000002) += (28, 1.35) -= (28, 1.65)
(29, 7.6499999999999995) += (29, 3.35) -= (29, 2.65)
(30, 9.350000000000001) += (30, 4.6499999999999995) -= (30, 5.35)
(31, 10.35) += (31, 6.6499999999999995) -= (31, 4.35)
};
\addplot [dashed, mark=none] coordinates {
   (-0.5,0)
   (-0.5,282.0-29.08607914449797)
   (0.5,282.0-29.08607914449797)
   (0.5,9.0-5.196152422706632)
   (1.5,9.0-5.196152422706632)
   (1.5,9.0-5.196152422706632)
   (2.5,9.0-5.196152422706632)
   (2.5,9.0-5.196152422706632)
   (3.5,9.0-5.196152422706632)
   (3.5,15.5-6.819090848492927)
   (4.5,15.5-6.819090848492927)
   (4.5,4.0-3.464101615137755)
   (5.5,4.0-3.464101615137755)
   (5.5,4.0-3.464101615137755)
   (6.5,4.0-3.464101615137755)
   (6.5,4.0-3.464101615137755)
   (7.5,4.0-3.464101615137755)
   (7.5,15.5-6.819090848492927)
   (8.5,15.5-6.819090848492927)
   (8.5,4.0-3.464101615137755)
   (9.5,4.0-3.464101615137755)
   (9.5,4.0-3.464101615137755)
   (10.5,4.0-3.464101615137755)
   (10.5,4.0-3.464101615137755)
   (11.5,4.0-3.464101615137755)
   (11.5,15.5-6.819090848492927)
   (12.5,15.5-6.819090848492927)
   (12.5,4.0-3.464101615137755)
   (13.5,4.0-3.464101615137755)
   (13.5,4.0-3.464101615137755)
   (14.5,4.0-3.464101615137755)
   (14.5,4.0-3.464101615137755)
   (15.5,4.0-3.464101615137755)
   (15.5,15.5-6.819090848492927)
   (16.5,15.5-6.819090848492927)
   (16.5,4.0-3.464101615137755)
   (17.5,4.0-3.464101615137755)
   (17.5,4.0-3.464101615137755)
   (18.5,4.0-3.464101615137755)
   (18.5,4.0-3.464101615137755)
   (19.5,4.0-3.464101615137755)
   (19.5,15.5-6.819090848492927)
   (20.5,15.5-6.819090848492927)
   (20.5,4.0-3.464101615137755)
   (21.5,4.0-3.464101615137755)
   (21.5,4.0-3.464101615137755)
   (22.5,4.0-3.464101615137755)
   (22.5,4.0-3.464101615137755)
   (23.5,4.0-3.464101615137755)
   (23.5,15.5-6.819090848492927)
   (24.5,15.5-6.819090848492927)
   (24.5,4.0-3.464101615137755)
   (25.5,4.0-3.464101615137755)
   (25.5,4.0-3.464101615137755)
   (26.5,4.0-3.464101615137755)
   (26.5,4.0-3.464101615137755)
   (27.5,4.0-3.464101615137755)
   (27.5,15.5-6.819090848492927)
   (28.5,15.5-6.819090848492927)
   (28.5,4.0-3.464101615137755)
   (29.5,4.0-3.464101615137755)
   (29.5,4.0-3.464101615137755)
   (30.5,4.0-3.464101615137755)
   (30.5,4.0-3.464101615137755)
   (31.5,4.0-3.464101615137755)
   (31.5,0)
};
\addlegendentry{5-Drzewker-1ac-pu-c}
\addlegendentry {3 RMS Error, 3$\times$500 shots}
\addlegendentry{Honeywell {\bf 25.6\%}} 
\end{axis}
\end{tikzpicture}
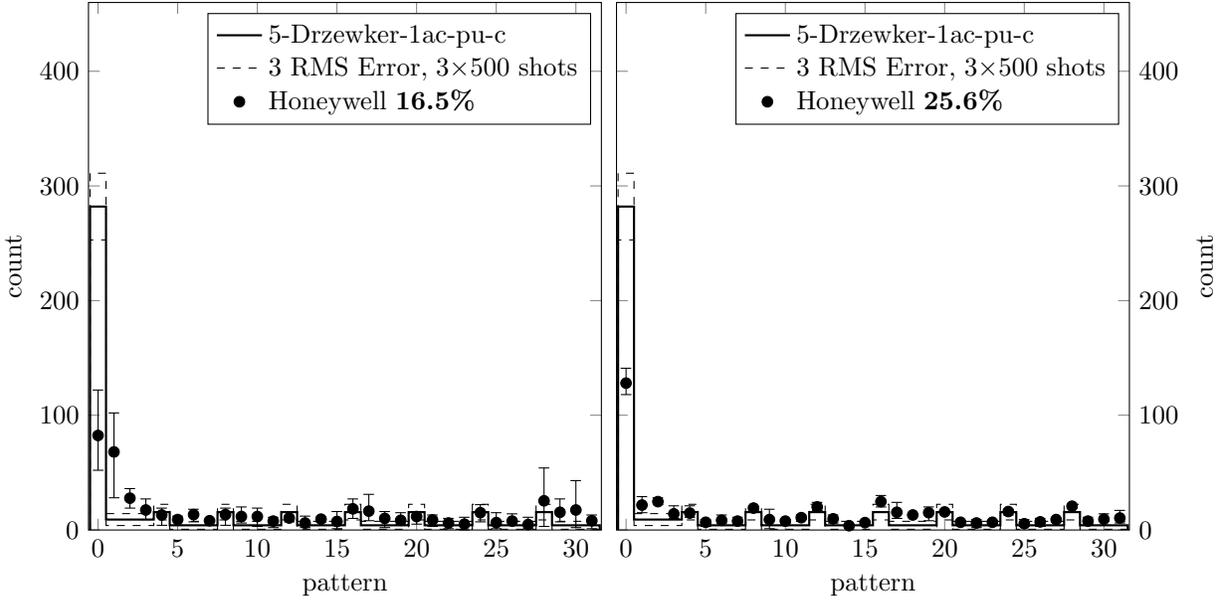
\caption{5-qubit 44CX Drzewker, 1 ancilla, partial uncompute: session 1 (left), session 2 (right).}
    \label{fig:5Drzewker}
\end{ffigure}

\begin{ffigure}
\begin{tikzpicture}
\begin{axis}[
    height=8.6cm,  
    width=8.4cm,
    xmin=-0.6,xmax=31.6,ymin=-0.1,ymax=660,
    legend cell align=left,
    ylabel ={count}, ylabel near ticks, yticklabel pos=left,
    xlabel={pattern},
    xlabel near ticks, xticklabel pos=bottom,
]
\addplot [thick, mark=none] coordinates {
   (-0.5,0)
   (-0.5,390.5)
   (0.5,390.5)
   (0.5,0.0)
   (1.5,0.0)
   (1.5,0.0)
   (2.5,0.0)
   (2.5,0.0)
   (3.5,0.0)
   (3.5,4.0)
   (4.5,4.0)
   (4.5,4.0)
   (5.5,4.0)
   (5.5,4.0)
   (6.5,4.0)
   (6.5,4.0)
   (7.5,4.0)
   (7.5,4.0)
   (8.5,4.0)
   (8.5,4.0)
   (9.5,4.0)
   (9.5,4.0)
   (10.5,4.0)
   (10.5,4.0)
   (11.5,4.0)
   (11.5,4.0)
   (12.5,4.0)
   (12.5,4.0)
   (13.5,4.0)
   (13.5,4.0)
   (14.5,4.0)
   (14.5,4.0)
   (15.5,4.0)
   (15.5,4.0)
   (16.5,4.0)
   (16.5,4.0)
   (17.5,4.0)
   (17.5,4.0)
   (18.5,4.0)
   (18.5,4.0)
   (19.5,4.0)
   (19.5,4.0)
   (20.5,4.0)
   (20.5,4.0)
   (21.5,4.0)
   (21.5,4.0)
   (22.5,4.0)
   (22.5,4.0)
   (23.5,4.0)
   (23.5,4.0)
   (24.5,4.0)
   (24.5,4.0)
   (25.5,4.0)
   (25.5,4.0)
   (26.5,4.0)
   (26.5,4.0)
   (27.5,4.0)
   (27.5,4.0)
   (28.5,4.0)
   (28.5,4.0)
   (29.5,4.0)
   (29.5,4.0)
   (30.5,4.0)
   (30.5,4.0)
   (31.5,4.0)
   (31.5,0)
};
\addplot [dashed, mark=none] coordinates { 
   (-0.5,0)
   (-0.5,390.5+34.22718218024966)
   (0.5,390.5+34.22718218024966)
   (0.5,0.0+0.0)
   (1.5,0.0+0.0)
   (1.5,0.0+0.0)
   (2.5,0.0+0.0)
   (2.5,0.0+0.0)
   (3.5,0.0+0.0)
   (3.5,4.0+3.464101615137755)
   (4.5,4.0+3.464101615137755)
   (4.5,4.0+3.464101615137755)
   (5.5,4.0+3.464101615137755)
   (5.5,4.0+3.464101615137755)
   (6.5,4.0+3.464101615137755)
   (6.5,4.0+3.464101615137755)
   (7.5,4.0+3.464101615137755)
   (7.5,4.0+3.464101615137755)
   (8.5,4.0+3.464101615137755)
   (8.5,4.0+3.464101615137755)
   (9.5,4.0+3.464101615137755)
   (9.5,4.0+3.464101615137755)
   (10.5,4.0+3.464101615137755)
   (10.5,4.0+3.464101615137755)
   (11.5,4.0+3.464101615137755)
   (11.5,4.0+3.464101615137755)
   (12.5,4.0+3.464101615137755)
   (12.5,4.0+3.464101615137755)
   (13.5,4.0+3.464101615137755)
   (13.5,4.0+3.464101615137755)
   (14.5,4.0+3.464101615137755)
   (14.5,4.0+3.464101615137755)
   (15.5,4.0+3.464101615137755)
   (15.5,4.0+3.464101615137755)
   (16.5,4.0+3.464101615137755)
   (16.5,4.0+3.464101615137755)
   (17.5,4.0+3.464101615137755)
   (17.5,4.0+3.464101615137755)
   (18.5,4.0+3.464101615137755)
   (18.5,4.0+3.464101615137755)
   (19.5,4.0+3.464101615137755)
   (19.5,4.0+3.464101615137755)
   (20.5,4.0+3.464101615137755)
   (20.5,4.0+3.464101615137755)
   (21.5,4.0+3.464101615137755)
   (21.5,4.0+3.464101615137755)
   (22.5,4.0+3.464101615137755)
   (22.5,4.0+3.464101615137755)
   (23.5,4.0+3.464101615137755)
   (23.5,4.0+3.464101615137755)
   (24.5,4.0+3.464101615137755)
   (24.5,4.0+3.464101615137755)
   (25.5,4.0+3.464101615137755)
   (25.5,4.0+3.464101615137755)
   (26.5,4.0+3.464101615137755)
   (26.5,4.0+3.464101615137755)
   (27.5,4.0+3.464101615137755)
   (27.5,4.0+3.464101615137755)
   (28.5,4.0+3.464101615137755)
   (28.5,4.0+3.464101615137755)
   (29.5,4.0+3.464101615137755)
   (29.5,4.0+3.464101615137755)
   (30.5,4.0+3.464101615137755)
   (30.5,4.0+3.464101615137755)
   (31.5,4.0+3.464101615137755)
   (31.5,0)
};
\addplot [only marks, error bars,y dir=both, y explicit] coordinates {
(0, 65.5) += (0, 59.5) -= (0, 44.55)
(1, 64.75) += (1, 89.25) -= (1, 53.75)
(2, 13.75) += (2, 6.25) -= (2, 5.75)
(3, 8.5) += (3, 17.5) -= (3, 8.5)
(4, 21.0) += (4, 13.0) -= (4, 13.0)
(5, 8.5) += (5, 4.5) -= (5, 2.5)
(6, 16.5) += (6, 8.5) -= (6, 13.5)
(7, 7.75) += (7, 3.25) -= (7, 2.75)
(8, 18.75) += (8, 10.25) -= (8, 18.75)
(9, 8.5) += (9, 6.5) -= (9, 3.5)
(10, 16.0) += (10, 6.949999999999999) -= (10, 15.0)
(11, 6.75) += (11, 6.25) -= (11, 3.75)
(12, 9.5) += (12, 7.45) -= (12, 8.5)
(13, 5.25) += (13, 5.75) -= (13, 5.25)
(14, 6.25) += (14, 9.75) -= (14, 4.25)
(15, 4.25) += (15, 4.75) -= (15, 4.25)
(16, 21.75) += (16, 13.25) -= (16, 11.75)
(17, 10.0) += (17, 6.0) -= (17, 4.0)
(18, 13.25) += (18, 9.75) -= (18, 7.25)
(19, 14.25) += (19, 11.75) -= (19, 8.25)
(20, 9.75) += (20, 3.25) -= (20, 5.75)
(21, 7.0) += (21, 5.0) -= (21, 3.0)
(22, 6.5) += (22, 4.5) -= (22, 6.5)
(23, 5.75) += (23, 6.25) -= (23, 4.75)
(24, 15.25) += (24, 19.75) -= (24, 11.25)
(25, 4.5) += (25, 4.5) -= (25, 3.5)
(26, 14.5) += (26, 12.5) -= (26, 6.5)
(27, 5.25) += (27, 1.75) -= (27, 3.25)
(28, 34.45) += (28, 33.550000000000004) -= (28, 31.45)
(29, 15.5) += (29, 10.5) -= (29, 11.5)
(30, 27.45) += (30, 34.4) -= (30, 23.45)
(31, 13.5) += (31, 10.45) -= (31, 12.5)
};
\addplot [dashed, mark=none] coordinates {
   (-0.5,0)
   (-0.5,390.5-34.22718218024966)
   (0.5,390.5-34.22718218024966)
   (0.5,0.0-0.0)
   (1.5,0.0-0.0)
   (1.5,0.0-0.0)
   (2.5,0.0-0.0)
   (2.5,0.0-0.0)
   (3.5,0.0-0.0)
   (3.5,4.0-3.464101615137755)
   (4.5,4.0-3.464101615137755)
   (4.5,4.0-3.464101615137755)
   (5.5,4.0-3.464101615137755)
   (5.5,4.0-3.464101615137755)
   (6.5,4.0-3.464101615137755)
   (6.5,4.0-3.464101615137755)
   (7.5,4.0-3.464101615137755)
   (7.5,4.0-3.464101615137755)
   (8.5,4.0-3.464101615137755)
   (8.5,4.0-3.464101615137755)
   (9.5,4.0-3.464101615137755)
   (9.5,4.0-3.464101615137755)
   (10.5,4.0-3.464101615137755)
   (10.5,4.0-3.464101615137755)
   (11.5,4.0-3.464101615137755)
   (11.5,4.0-3.464101615137755)
   (12.5,4.0-3.464101615137755)
   (12.5,4.0-3.464101615137755)
   (13.5,4.0-3.464101615137755)
   (13.5,4.0-3.464101615137755)
   (14.5,4.0-3.464101615137755)
   (14.5,4.0-3.464101615137755)
   (15.5,4.0-3.464101615137755)
   (15.5,4.0-3.464101615137755)
   (16.5,4.0-3.464101615137755)
   (16.5,4.0-3.464101615137755)
   (17.5,4.0-3.464101615137755)
   (17.5,4.0-3.464101615137755)
   (18.5,4.0-3.464101615137755)
   (18.5,4.0-3.464101615137755)
   (19.5,4.0-3.464101615137755)
   (19.5,4.0-3.464101615137755)
   (20.5,4.0-3.464101615137755)
   (20.5,4.0-3.464101615137755)
   (21.5,4.0-3.464101615137755)
   (21.5,4.0-3.464101615137755)
   (22.5,4.0-3.464101615137755)
   (22.5,4.0-3.464101615137755)
   (23.5,4.0-3.464101615137755)
   (23.5,4.0-3.464101615137755)
   (24.5,4.0-3.464101615137755)
   (24.5,4.0-3.464101615137755)
   (25.5,4.0-3.464101615137755)
   (25.5,4.0-3.464101615137755)
   (26.5,4.0-3.464101615137755)
   (26.5,4.0-3.464101615137755)
   (27.5,4.0-3.464101615137755)
   (27.5,4.0-3.464101615137755)
   (28.5,4.0-3.464101615137755)
   (28.5,4.0-3.464101615137755)
   (29.5,4.0-3.464101615137755)
   (29.5,4.0-3.464101615137755)
   (30.5,4.0-3.464101615137755)
   (30.5,4.0-3.464101615137755)
   (31.5,4.0-3.464101615137755)
   (31.5,0)
};
\addlegendentry{5-Wojter-1ac-pu-a}
\addlegendentry {3 RMS Error, 3$\times$500 shots}
\addlegendentry{Honeywell {\bf 13.1\%}} 
\end{axis}
\end{tikzpicture}
\begin{tikzpicture}
\begin{axis}[
    height=8.6cm,  
    width=8.4cm,
    xmin=-0.6,xmax=31.6,ymin=-0.1,ymax=660,
    legend cell align=left,
    ylabel ={count}, ylabel near ticks, yticklabel pos=right,
    xlabel={pattern},
    xlabel near ticks, xticklabel pos=bottom,
]
\addplot [thick, mark=none] coordinates {
   (-0.5,0)
   (-0.5,483.5)
   (0.5,483.5)
   (0.5,5.0)
   (1.5,5.0)
   (1.5,5.0)
   (2.5,5.0)
   (2.5,5.0)
   (3.5,5.0)
   (3.5,0.0)
   (4.5,0.0)
   (4.5,0.0)
   (5.5,0.0)
   (5.5,0.0)
   (6.5,0.0)
   (6.5,0.0)
   (7.5,0.0)
   (7.5,0.0)
   (8.5,0.0)
   (8.5,0.0)
   (9.5,0.0)
   (9.5,0.0)
   (10.5,0.0)
   (10.5,0.0)
   (11.5,0.0)
   (11.5,0.0)
   (12.5,0.0)
   (12.5,0.0)
   (13.5,0.0)
   (13.5,0.0)
   (14.5,0.0)
   (14.5,0.0)
   (15.5,0.0)
   (15.5,0.0)
   (16.5,0.0)
   (16.5,0.0)
   (17.5,0.0)
   (17.5,0.0)
   (18.5,0.0)
   (18.5,0.0)
   (19.5,0.0)
   (19.5,0.0)
   (20.5,0.0)
   (20.5,0.0)
   (21.5,0.0)
   (21.5,0.0)
   (22.5,0.0)
   (22.5,0.0)
   (23.5,0.0)
   (23.5,0.0)
   (24.5,0.0)
   (24.5,0.0)
   (25.5,0.0)
   (25.5,0.0)
   (26.5,0.0)
   (26.5,0.0)
   (27.5,0.0)
   (27.5,0.0)
   (28.5,0.0)
   (28.5,0.0)
   (29.5,0.0)
   (29.5,0.0)
   (30.5,0.0)
   (30.5,0.0)
   (31.5,0.0)
   (31.5,0)
};
\addplot [dashed, mark=none] coordinates { 
   (-0.5,0)
   (-0.5,483.5+38.08543028508409)
   (0.5,483.5+38.08543028508409)
   (0.5,5.0+3.872983346207417)
   (1.5,5.0+3.872983346207417)
   (1.5,5.0+3.872983346207417)
   (2.5,5.0+3.872983346207417)
   (2.5,5.0+3.872983346207417)
   (3.5,5.0+3.872983346207417)
   (3.5,0.0+0.0)
   (4.5,0.0+0.0)
   (4.5,0.0+0.0)
   (5.5,0.0+0.0)
   (5.5,0.0+0.0)
   (6.5,0.0+0.0)
   (6.5,0.0+0.0)
   (7.5,0.0+0.0)
   (7.5,0.0+0.0)
   (8.5,0.0+0.0)
   (8.5,0.0+0.0)
   (9.5,0.0+0.0)
   (9.5,0.0+0.0)
   (10.5,0.0+0.0)
   (10.5,0.0+0.0)
   (11.5,0.0+0.0)
   (11.5,0.0+0.0)
   (12.5,0.0+0.0)
   (12.5,0.0+0.0)
   (13.5,0.0+0.0)
   (13.5,0.0+0.0)
   (14.5,0.0+0.0)
   (14.5,0.0+0.0)
   (15.5,0.0+0.0)
   (15.5,0.0+0.0)
   (16.5,0.0+0.0)
   (16.5,0.0+0.0)
   (17.5,0.0+0.0)
   (17.5,0.0+0.0)
   (18.5,0.0+0.0)
   (18.5,0.0+0.0)
   (19.5,0.0+0.0)
   (19.5,0.0+0.0)
   (20.5,0.0+0.0)
   (20.5,0.0+0.0)
   (21.5,0.0+0.0)
   (21.5,0.0+0.0)
   (22.5,0.0+0.0)
   (22.5,0.0+0.0)
   (23.5,0.0+0.0)
   (23.5,0.0+0.0)
   (24.5,0.0+0.0)
   (24.5,0.0+0.0)
   (25.5,0.0+0.0)
   (25.5,0.0+0.0)
   (26.5,0.0+0.0)
   (26.5,0.0+0.0)
   (27.5,0.0+0.0)
   (27.5,0.0+0.0)
   (28.5,0.0+0.0)
   (28.5,0.0+0.0)
   (29.5,0.0+0.0)
   (29.5,0.0+0.0)
   (30.5,0.0+0.0)
   (30.5,0.0+0.0)
   (31.5,0.0+0.0)
   (31.5,0)
};
\addplot [only marks, error bars,y dir=both, y explicit] coordinates {
(0, 62.0) += (0, 46.0) -= (0, 44.0)
(1, 18.0) += (1, 5.0) -= (1, 3.0)
(2, 30.65) += (2, 32.349999999999994) -= (2, 16.650000000000002)
(3, 17.0) += (3, 5.0) -= (3, 6.0)
(4, 13.65) += (4, 12.35) -= (4, 9.65)
(5, 10.0) += (5, 1.0) -= (5, 1.0)
(6, 11.350000000000001) += (6, 1.65) -= (6, 2.35)
(7, 9.0) += (7, 2.0) -= (7, 3.0)
(8, 12.0) += (8, 6.0) -= (8, 4.0)
(9, 10.0) += (9, 8.0) -= (9, 5.0)
(10, 11.65) += (10, 4.35) -= (10, 2.65)
(11, 13.0) += (11, 9.0) -= (11, 6.0)
(12, 15.0) += (12, 2.0) -= (12, 3.0)
(13, 10.35) += (13, 2.65) -= (13, 1.35)
(14, 7.0) += (14, 1.0) -= (14, 1.0)
(15, 11.350000000000001) += (15, 5.6499999999999995) -= (15, 7.35)
(16, 18.0) += (16, 12.0) -= (16, 10.0)
(17, 15.65) += (17, 3.35) -= (17, 2.65)
(18, 19.650000000000002) += (18, 13.350000000000001) -= (18, 8.65)
(19, 15.0) += (19, 3.0) -= (19, 2.0)
(20, 12.0) += (20, 1.0) -= (20, 2.0)
(21, 7.35) += (21, 1.65) -= (21, 1.35)
(22, 9.350000000000001) += (22, 3.65) -= (22, 3.35)
(23, 9.350000000000001) += (23, 1.65) -= (23, 1.35)
(24, 11.350000000000001) += (24, 4.6499999999999995) -= (24, 4.35)
(25, 11.0) += (25, 7.0) -= (25, 4.0)
(26, 8.35) += (26, 1.65) -= (26, 1.35)
(27, 15.350000000000001) += (27, 2.65) -= (27, 2.35)
(28, 26.65) += (28, 21.35) -= (28, 14.65)
(29, 18.0) += (29, 10.0) -= (29, 5.0)
(30, 16.35) += (30, 7.6499999999999995) -= (30, 11.350000000000001)
(31, 24.65) += (31, 14.35) -= (31, 10.65)
};
\addplot [dashed, mark=none] coordinates {
   (-0.5,0)
   (-0.5,483.5-38.08543028508409)
   (0.5,483.5-38.08543028508409)
   (0.5,5.0-3.872983346207417)
   (1.5,5.0-3.872983346207417)
   (1.5,5.0-3.872983346207417)
   (2.5,5.0-3.872983346207417)
   (2.5,5.0-3.872983346207417)
   (3.5,5.0-3.872983346207417)
   (3.5,0.0-0.0)
   (4.5,0.0-0.0)
   (4.5,0.0-0.0)
   (5.5,0.0-0.0)
   (5.5,0.0-0.0)
   (6.5,0.0-0.0)
   (6.5,0.0-0.0)
   (7.5,0.0-0.0)
   (7.5,0.0-0.0)
   (8.5,0.0-0.0)
   (8.5,0.0-0.0)
   (9.5,0.0-0.0)
   (9.5,0.0-0.0)
   (10.5,0.0-0.0)
   (10.5,0.0-0.0)
   (11.5,0.0-0.0)
   (11.5,0.0-0.0)
   (12.5,0.0-0.0)
   (12.5,0.0-0.0)
   (13.5,0.0-0.0)
   (13.5,0.0-0.0)
   (14.5,0.0-0.0)
   (14.5,0.0-0.0)
   (15.5,0.0-0.0)
   (15.5,0.0-0.0)
   (16.5,0.0-0.0)
   (16.5,0.0-0.0)
   (17.5,0.0-0.0)
   (17.5,0.0-0.0)
   (18.5,0.0-0.0)
   (18.5,0.0-0.0)
   (19.5,0.0-0.0)
   (19.5,0.0-0.0)
   (20.5,0.0-0.0)
   (20.5,0.0-0.0)
   (21.5,0.0-0.0)
   (21.5,0.0-0.0)
   (22.5,0.0-0.0)
   (22.5,0.0-0.0)
   (23.5,0.0-0.0)
   (23.5,0.0-0.0)
   (24.5,0.0-0.0)
   (24.5,0.0-0.0)
   (25.5,0.0-0.0)
   (25.5,0.0-0.0)
   (26.5,0.0-0.0)
   (26.5,0.0-0.0)
   (27.5,0.0-0.0)
   (27.5,0.0-0.0)
   (28.5,0.0-0.0)
   (28.5,0.0-0.0)
   (29.5,0.0-0.0)
   (29.5,0.0-0.0)
   (30.5,0.0-0.0)
   (30.5,0.0-0.0)
   (31.5,0.0-0.0)
   (31.5,0)
};
\addlegendentry{5-WojterAA-1ac-pu}
\addlegendentry {3 RMS Error, 3$\times$500 shots}
\addlegendentry{Honeywell {\bf 12.4\%}} 
\end{axis}
\end{tikzpicture}
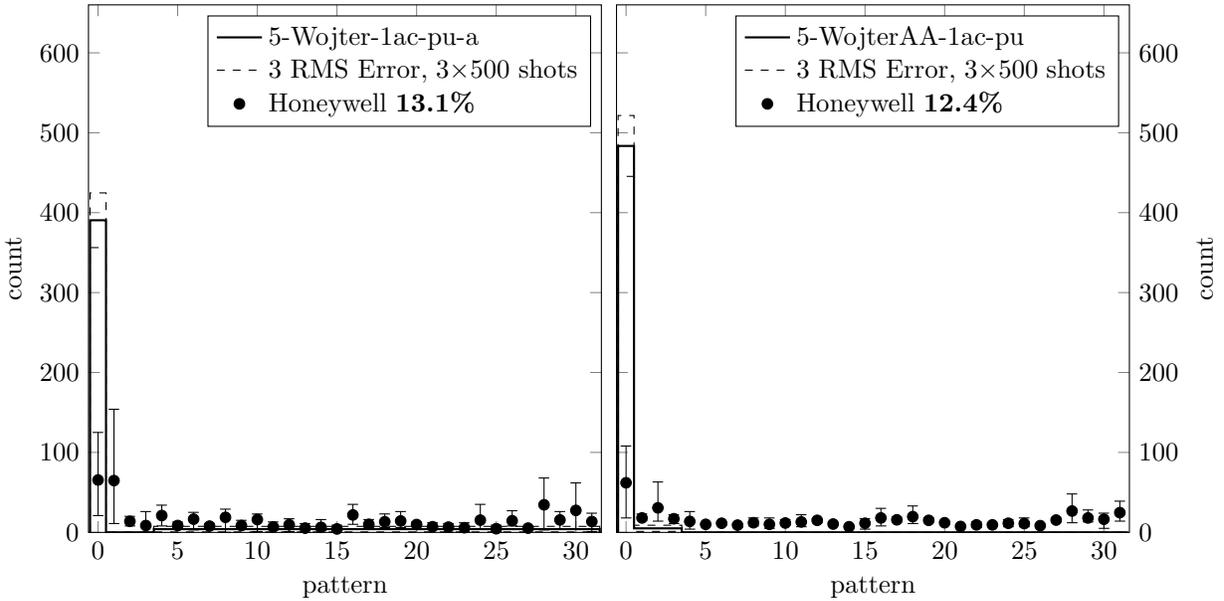
\caption{5-qubit Wojter with 1 ancilla partial uncompute: plain 31CX (left), variant utilizing additional oracle and full Grover diffuser at the end (right).}
\end{ffigure}

\begin{ffigure}
\begin{tikzpicture}
\begin{axis}[
    height=8.6cm,  
    width=8.4cm,
    xmin=-0.6,xmax=31.6,ymin=-0.1,ymax=260,
    legend cell align=left,
    ylabel ={count}, ylabel near ticks, yticklabel pos=left,
    xlabel={pattern},
    xlabel near ticks, xticklabel pos=bottom,
]
\addplot [thick, mark=none] coordinates {
   (-0.5,0)
   (-0.5,118.0)
   (0.5,118.0)
   (0.5,15.5)
   (1.5,15.5)
   (1.5,9.0)
   (2.5,9.0)
   (2.5,15.5)
   (3.5,15.5)
   (3.5,9.0)
   (4.5,9.0)
   (4.5,15.5)
   (5.5,15.5)
   (5.5,9.0)
   (6.5,9.0)
   (6.5,15.5)
   (7.5,15.5)
   (7.5,9.0)
   (8.5,9.0)
   (8.5,15.5)
   (9.5,15.5)
   (9.5,9.0)
   (10.5,9.0)
   (10.5,15.5)
   (11.5,15.5)
   (11.5,9.0)
   (12.5,9.0)
   (12.5,15.5)
   (13.5,15.5)
   (13.5,9.0)
   (14.5,9.0)
   (14.5,15.5)
   (15.5,15.5)
   (15.5,9.0)
   (16.5,9.0)
   (16.5,15.5)
   (17.5,15.5)
   (17.5,9.0)
   (18.5,9.0)
   (18.5,15.5)
   (19.5,15.5)
   (19.5,9.0)
   (20.5,9.0)
   (20.5,15.5)
   (21.5,15.5)
   (21.5,9.0)
   (22.5,9.0)
   (22.5,15.5)
   (23.5,15.5)
   (23.5,9.0)
   (24.5,9.0)
   (24.5,15.5)
   (25.5,15.5)
   (25.5,9.0)
   (26.5,9.0)
   (26.5,15.5)
   (27.5,15.5)
   (27.5,9.0)
   (28.5,9.0)
   (28.5,15.5)
   (29.5,15.5)
   (29.5,9.0)
   (30.5,9.0)
   (30.5,15.5)
   (31.5,15.5)
   (31.5,0)
};
\addplot [dashed, mark=none] coordinates { 
   (-0.5,0)
   (-0.5,118.0+18.81488772222678)
   (0.5,118.0+18.81488772222678)
   (0.5,15.5+6.819090848492927)
   (1.5,15.5+6.819090848492927)
   (1.5,9.0+5.196152422706632)
   (2.5,9.0+5.196152422706632)
   (2.5,15.5+6.819090848492927)
   (3.5,15.5+6.819090848492927)
   (3.5,9.0+5.196152422706632)
   (4.5,9.0+5.196152422706632)
   (4.5,15.5+6.819090848492927)
   (5.5,15.5+6.819090848492927)
   (5.5,9.0+5.196152422706632)
   (6.5,9.0+5.196152422706632)
   (6.5,15.5+6.819090848492927)
   (7.5,15.5+6.819090848492927)
   (7.5,9.0+5.196152422706632)
   (8.5,9.0+5.196152422706632)
   (8.5,15.5+6.819090848492927)
   (9.5,15.5+6.819090848492927)
   (9.5,9.0+5.196152422706632)
   (10.5,9.0+5.196152422706632)
   (10.5,15.5+6.819090848492927)
   (11.5,15.5+6.819090848492927)
   (11.5,9.0+5.196152422706632)
   (12.5,9.0+5.196152422706632)
   (12.5,15.5+6.819090848492927)
   (13.5,15.5+6.819090848492927)
   (13.5,9.0+5.196152422706632)
   (14.5,9.0+5.196152422706632)
   (14.5,15.5+6.819090848492927)
   (15.5,15.5+6.819090848492927)
   (15.5,9.0+5.196152422706632)
   (16.5,9.0+5.196152422706632)
   (16.5,15.5+6.819090848492927)
   (17.5,15.5+6.819090848492927)
   (17.5,9.0+5.196152422706632)
   (18.5,9.0+5.196152422706632)
   (18.5,15.5+6.819090848492927)
   (19.5,15.5+6.819090848492927)
   (19.5,9.0+5.196152422706632)
   (20.5,9.0+5.196152422706632)
   (20.5,15.5+6.819090848492927)
   (21.5,15.5+6.819090848492927)
   (21.5,9.0+5.196152422706632)
   (22.5,9.0+5.196152422706632)
   (22.5,15.5+6.819090848492927)
   (23.5,15.5+6.819090848492927)
   (23.5,9.0+5.196152422706632)
   (24.5,9.0+5.196152422706632)
   (24.5,15.5+6.819090848492927)
   (25.5,15.5+6.819090848492927)
   (25.5,9.0+5.196152422706632)
   (26.5,9.0+5.196152422706632)
   (26.5,15.5+6.819090848492927)
   (27.5,15.5+6.819090848492927)
   (27.5,9.0+5.196152422706632)
   (28.5,9.0+5.196152422706632)
   (28.5,15.5+6.819090848492927)
   (29.5,15.5+6.819090848492927)
   (29.5,9.0+5.196152422706632)
   (30.5,9.0+5.196152422706632)
   (30.5,15.5+6.819090848492927)
   (31.5,15.5+6.819090848492927)
   (31.5,0)
};
\addplot [only marks, error bars,y dir=both, y explicit] coordinates {
(0, 78.0) += (0, 11.0) -= (0, 14.0)
(1, 32.65) += (1, 14.35) -= (1, 9.65)
(2, 10.35) += (2, 4.6499999999999995) -= (2, 4.35)
(3, 9.350000000000001) += (3, 3.65) -= (3, 4.35)
(4, 9.65) += (4, 3.35) -= (4, 3.65)
(5, 9.350000000000001) += (5, 3.65) -= (5, 4.35)
(6, 10.65) += (6, 3.35) -= (6, 2.65)
(7, 15.65) += (7, 7.35) -= (7, 8.65)
(8, 9.0) += (8, 2.0) -= (8, 1.0)
(9, 15.0) += (9, 3.0) -= (9, 3.0)
(10, 7.6499999999999995) += (10, 4.35) -= (10, 5.6499999999999995)
(11, 13.0) += (11, 8.0) -= (11, 8.0)
(12, 9.0) += (12, 2.0) -= (12, 2.0)
(13, 10.65) += (13, 4.35) -= (13, 3.65)
(14, 13.350000000000001) += (14, 9.65) -= (14, 5.35)
(15, 18.0) += (15, 4.0) -= (15, 6.0)
(16, 11.350000000000001) += (16, 4.6499999999999995) -= (16, 3.35)
(17, 14.0) += (17, 2.0) -= (17, 4.0)
(18, 8.65) += (18, 2.35) -= (18, 1.65)
(19, 10.35) += (19, 3.65) -= (19, 2.35)
(20, 5.6499999999999995) += (20, 4.35) -= (20, 2.65)
(21, 8.65) += (21, 1.35) -= (21, 1.65)
(22, 16.650000000000002) += (22, 4.35) -= (22, 2.65)
(23, 17.65) += (23, 9.350000000000001) -= (23, 4.6499999999999995)
(24, 8.35) += (24, 1.65) -= (24, 1.35)
(25, 9.0) += (25, 4.0) -= (25, 3.0)
(26, 11.65) += (26, 2.35) -= (26, 3.65)
(27, 20.650000000000002) += (27, 5.35) -= (27, 7.6499999999999995)
(28, 12.0) += (28, 4.0) -= (28, 4.0)
(29, 15.0) += (29, 8.0) -= (29, 4.0)
(30, 28.65) += (30, 4.35) -= (30, 2.65)
(31, 30.349999999999998) += (31, 1.65) -= (31, 1.35)
};
\addplot [dashed, mark=none] coordinates {
   (-0.5,0)
   (-0.5,118.0-18.81488772222678)
   (0.5,118.0-18.81488772222678)
   (0.5,15.5-6.819090848492927)
   (1.5,15.5-6.819090848492927)
   (1.5,9.0-5.196152422706632)
   (2.5,9.0-5.196152422706632)
   (2.5,15.5-6.819090848492927)
   (3.5,15.5-6.819090848492927)
   (3.5,9.0-5.196152422706632)
   (4.5,9.0-5.196152422706632)
   (4.5,15.5-6.819090848492927)
   (5.5,15.5-6.819090848492927)
   (5.5,9.0-5.196152422706632)
   (6.5,9.0-5.196152422706632)
   (6.5,15.5-6.819090848492927)
   (7.5,15.5-6.819090848492927)
   (7.5,9.0-5.196152422706632)
   (8.5,9.0-5.196152422706632)
   (8.5,15.5-6.819090848492927)
   (9.5,15.5-6.819090848492927)
   (9.5,9.0-5.196152422706632)
   (10.5,9.0-5.196152422706632)
   (10.5,15.5-6.819090848492927)
   (11.5,15.5-6.819090848492927)
   (11.5,9.0-5.196152422706632)
   (12.5,9.0-5.196152422706632)
   (12.5,15.5-6.819090848492927)
   (13.5,15.5-6.819090848492927)
   (13.5,9.0-5.196152422706632)
   (14.5,9.0-5.196152422706632)
   (14.5,15.5-6.819090848492927)
   (15.5,15.5-6.819090848492927)
   (15.5,9.0-5.196152422706632)
   (16.5,9.0-5.196152422706632)
   (16.5,15.5-6.819090848492927)
   (17.5,15.5-6.819090848492927)
   (17.5,9.0-5.196152422706632)
   (18.5,9.0-5.196152422706632)
   (18.5,15.5-6.819090848492927)
   (19.5,15.5-6.819090848492927)
   (19.5,9.0-5.196152422706632)
   (20.5,9.0-5.196152422706632)
   (20.5,15.5-6.819090848492927)
   (21.5,15.5-6.819090848492927)
   (21.5,9.0-5.196152422706632)
   (22.5,9.0-5.196152422706632)
   (22.5,15.5-6.819090848492927)
   (23.5,15.5-6.819090848492927)
   (23.5,9.0-5.196152422706632)
   (24.5,9.0-5.196152422706632)
   (24.5,15.5-6.819090848492927)
   (25.5,15.5-6.819090848492927)
   (25.5,9.0-5.196152422706632)
   (26.5,9.0-5.196152422706632)
   (26.5,15.5-6.819090848492927)
   (27.5,15.5-6.819090848492927)
   (27.5,9.0-5.196152422706632)
   (28.5,9.0-5.196152422706632)
   (28.5,15.5-6.819090848492927)
   (29.5,15.5-6.819090848492927)
   (29.5,9.0-5.196152422706632)
   (30.5,9.0-5.196152422706632)
   (30.5,15.5-6.819090848492927)
   (31.5,15.5-6.819090848492927)
   (31.5,0)
};
\addlegendentry{5-Partial-1ac-a}
\addlegendentry {3 RMS Error, 3$\times$500 shots}
\addlegendentry{Honeywell {\bf 15.6\%}} 
\end{axis}
\end{tikzpicture}
\begin{tikzpicture}
\begin{axis}[
    height=8.6cm,  
    width=8.4cm,
    xmin=-0.6,xmax=31.6,ymin=-0.1,ymax=260,
    legend cell align=left,
    ylabel ={count}, ylabel near ticks, yticklabel pos=right,
    xlabel={pattern},
    xlabel near ticks, xticklabel pos=bottom,
]
\addplot [thick, mark=none] coordinates {
   (-0.5,0)
   (-0.5,97.5)
   (0.5,97.5)
   (0.5,15.5)
   (1.5,15.5)
   (1.5,15.5)
   (2.5,15.5)
   (2.5,15.5)
   (3.5,15.5)
   (3.5,4.0)
   (4.5,4.0)
   (4.5,15.5)
   (5.5,15.5)
   (5.5,15.5)
   (6.5,15.5)
   (6.5,15.5)
   (7.5,15.5)
   (7.5,4.0)
   (8.5,4.0)
   (8.5,15.5)
   (9.5,15.5)
   (9.5,15.5)
   (10.5,15.5)
   (10.5,15.5)
   (11.5,15.5)
   (11.5,4.0)
   (12.5,4.0)
   (12.5,15.5)
   (13.5,15.5)
   (13.5,15.5)
   (14.5,15.5)
   (14.5,15.5)
   (15.5,15.5)
   (15.5,4.0)
   (16.5,4.0)
   (16.5,15.5)
   (17.5,15.5)
   (17.5,15.5)
   (18.5,15.5)
   (18.5,15.5)
   (19.5,15.5)
   (19.5,4.0)
   (20.5,4.0)
   (20.5,15.5)
   (21.5,15.5)
   (21.5,15.5)
   (22.5,15.5)
   (22.5,15.5)
   (23.5,15.5)
   (23.5,4.0)
   (24.5,4.0)
   (24.5,15.5)
   (25.5,15.5)
   (25.5,15.5)
   (26.5,15.5)
   (26.5,15.5)
   (27.5,15.5)
   (27.5,4.0)
   (28.5,4.0)
   (28.5,15.5)
   (29.5,15.5)
   (29.5,15.5)
   (30.5,15.5)
   (30.5,15.5)
   (31.5,15.5)
   (31.5,0)
};
\addplot [dashed, mark=none] coordinates { 
   (-0.5,0)
   (-0.5,97.5+17.10263137648707)
   (0.5,97.5+17.10263137648707)
   (0.5,15.5+6.819090848492927)
   (1.5,15.5+6.819090848492927)
   (1.5,15.5+6.819090848492927)
   (2.5,15.5+6.819090848492927)
   (2.5,15.5+6.819090848492927)
   (3.5,15.5+6.819090848492927)
   (3.5,4.0+3.464101615137755)
   (4.5,4.0+3.464101615137755)
   (4.5,15.5+6.819090848492927)
   (5.5,15.5+6.819090848492927)
   (5.5,15.5+6.819090848492927)
   (6.5,15.5+6.819090848492927)
   (6.5,15.5+6.819090848492927)
   (7.5,15.5+6.819090848492927)
   (7.5,4.0+3.464101615137755)
   (8.5,4.0+3.464101615137755)
   (8.5,15.5+6.819090848492927)
   (9.5,15.5+6.819090848492927)
   (9.5,15.5+6.819090848492927)
   (10.5,15.5+6.819090848492927)
   (10.5,15.5+6.819090848492927)
   (11.5,15.5+6.819090848492927)
   (11.5,4.0+3.464101615137755)
   (12.5,4.0+3.464101615137755)
   (12.5,15.5+6.819090848492927)
   (13.5,15.5+6.819090848492927)
   (13.5,15.5+6.819090848492927)
   (14.5,15.5+6.819090848492927)
   (14.5,15.5+6.819090848492927)
   (15.5,15.5+6.819090848492927)
   (15.5,4.0+3.464101615137755)
   (16.5,4.0+3.464101615137755)
   (16.5,15.5+6.819090848492927)
   (17.5,15.5+6.819090848492927)
   (17.5,15.5+6.819090848492927)
   (18.5,15.5+6.819090848492927)
   (18.5,15.5+6.819090848492927)
   (19.5,15.5+6.819090848492927)
   (19.5,4.0+3.464101615137755)
   (20.5,4.0+3.464101615137755)
   (20.5,15.5+6.819090848492927)
   (21.5,15.5+6.819090848492927)
   (21.5,15.5+6.819090848492927)
   (22.5,15.5+6.819090848492927)
   (22.5,15.5+6.819090848492927)
   (23.5,15.5+6.819090848492927)
   (23.5,4.0+3.464101615137755)
   (24.5,4.0+3.464101615137755)
   (24.5,15.5+6.819090848492927)
   (25.5,15.5+6.819090848492927)
   (25.5,15.5+6.819090848492927)
   (26.5,15.5+6.819090848492927)
   (26.5,15.5+6.819090848492927)
   (27.5,15.5+6.819090848492927)
   (27.5,4.0+3.464101615137755)
   (28.5,4.0+3.464101615137755)
   (28.5,15.5+6.819090848492927)
   (29.5,15.5+6.819090848492927)
   (29.5,15.5+6.819090848492927)
   (30.5,15.5+6.819090848492927)
   (30.5,15.5+6.819090848492927)
   (31.5,15.5+6.819090848492927)
   (31.5,0)
};
\addplot [only marks, error bars,y dir=both, y explicit] coordinates {
(0, 66.0) += (0, 11.0) -= (0, 14.0)
(1, 20.35) += (1, 4.6499999999999995) -= (1, 5.35)
(2, 23.349999999999998) += (2, 4.6499999999999995) -= (2, 6.35)
(3, 28.0) += (3, 8.0) -= (3, 4.0)
(4, 10.35) += (4, 1.65) -= (4, 1.35)
(5, 10.35) += (5, 1.65) -= (5, 1.35)
(6, 10.0) += (6, 4.0) -= (6, 3.0)
(7, 16.0) += (7, 9.0) -= (7, 8.0)
(8, 11.350000000000001) += (8, 3.65) -= (8, 3.35)
(9, 9.65) += (9, 2.35) -= (9, 1.65)
(10, 8.35) += (10, 7.6499999999999995) -= (10, 4.35)
(11, 13.0) += (11, 1.0) -= (11, 1.0)
(12, 6.35) += (12, 0.65) -= (12, 1.35)
(13, 11.65) += (13, 3.35) -= (13, 3.65)
(14, 11.0) += (14, 3.0) -= (14, 2.0)
(15, 18.0) += (15, 1.0) -= (15, 2.0)
(16, 13.0) += (16, 2.0) -= (16, 4.0)
(17, 12.35) += (17, 2.65) -= (17, 2.35)
(18, 15.350000000000001) += (18, 5.6499999999999995) -= (18, 5.35)
(19, 15.0) += (19, 4.0) -= (19, 4.0)
(20, 8.65) += (20, 2.35) -= (20, 1.65)
(21, 12.0) += (21, 3.0) -= (21, 3.0)
(22, 9.350000000000001) += (22, 1.65) -= (22, 1.35)
(23, 10.65) += (23, 4.35) -= (23, 2.65)
(24, 8.35) += (24, 2.65) -= (24, 2.35)
(25, 11.350000000000001) += (25, 0.65) -= (25, 1.35)
(26, 10.65) += (26, 1.35) -= (26, 1.65)
(27, 13.350000000000001) += (27, 3.65) -= (27, 2.35)
(28, 12.35) += (28, 2.65) -= (28, 2.35)
(29, 22.65) += (29, 1.35) -= (29, 0.65)
(30, 22.0) += (30, 2.0) -= (30, 2.0)
(31, 29.35) += (31, 4.6499999999999995) -= (31, 6.35)
};
\addplot [dashed, mark=none] coordinates {
   (-0.5,0)
   (-0.5,97.5-17.10263137648707)
   (0.5,97.5-17.10263137648707)
   (0.5,15.5-6.819090848492927)
   (1.5,15.5-6.819090848492927)
   (1.5,15.5-6.819090848492927)
   (2.5,15.5-6.819090848492927)
   (2.5,15.5-6.819090848492927)
   (3.5,15.5-6.819090848492927)
   (3.5,4.0-3.464101615137755)
   (4.5,4.0-3.464101615137755)
   (4.5,15.5-6.819090848492927)
   (5.5,15.5-6.819090848492927)
   (5.5,15.5-6.819090848492927)
   (6.5,15.5-6.819090848492927)
   (6.5,15.5-6.819090848492927)
   (7.5,15.5-6.819090848492927)
   (7.5,4.0-3.464101615137755)
   (8.5,4.0-3.464101615137755)
   (8.5,15.5-6.819090848492927)
   (9.5,15.5-6.819090848492927)
   (9.5,15.5-6.819090848492927)
   (10.5,15.5-6.819090848492927)
   (10.5,15.5-6.819090848492927)
   (11.5,15.5-6.819090848492927)
   (11.5,4.0-3.464101615137755)
   (12.5,4.0-3.464101615137755)
   (12.5,15.5-6.819090848492927)
   (13.5,15.5-6.819090848492927)
   (13.5,15.5-6.819090848492927)
   (14.5,15.5-6.819090848492927)
   (14.5,15.5-6.819090848492927)
   (15.5,15.5-6.819090848492927)
   (15.5,4.0-3.464101615137755)
   (16.5,4.0-3.464101615137755)
   (16.5,15.5-6.819090848492927)
   (17.5,15.5-6.819090848492927)
   (17.5,15.5-6.819090848492927)
   (18.5,15.5-6.819090848492927)
   (18.5,15.5-6.819090848492927)
   (19.5,15.5-6.819090848492927)
   (19.5,4.0-3.464101615137755)
   (20.5,4.0-3.464101615137755)
   (20.5,15.5-6.819090848492927)
   (21.5,15.5-6.819090848492927)
   (21.5,15.5-6.819090848492927)
   (22.5,15.5-6.819090848492927)
   (22.5,15.5-6.819090848492927)
   (23.5,15.5-6.819090848492927)
   (23.5,4.0-3.464101615137755)
   (24.5,4.0-3.464101615137755)
   (24.5,15.5-6.819090848492927)
   (25.5,15.5-6.819090848492927)
   (25.5,15.5-6.819090848492927)
   (26.5,15.5-6.819090848492927)
   (26.5,15.5-6.819090848492927)
   (27.5,15.5-6.819090848492927)
   (27.5,4.0-3.464101615137755)
   (28.5,4.0-3.464101615137755)
   (28.5,15.5-6.819090848492927)
   (29.5,15.5-6.819090848492927)
   (29.5,15.5-6.819090848492927)
   (30.5,15.5-6.819090848492927)
   (30.5,15.5-6.819090848492927)
   (31.5,15.5-6.819090848492927)
   (31.5,0)
};
\addlegendentry{5-Partial-1ac-b}
\addlegendentry {3 RMS Error, 3$\times$500 shots}
\addlegendentry{Honeywell {\bf 13.2\%}} 
\end{axis}
\end{tikzpicture}
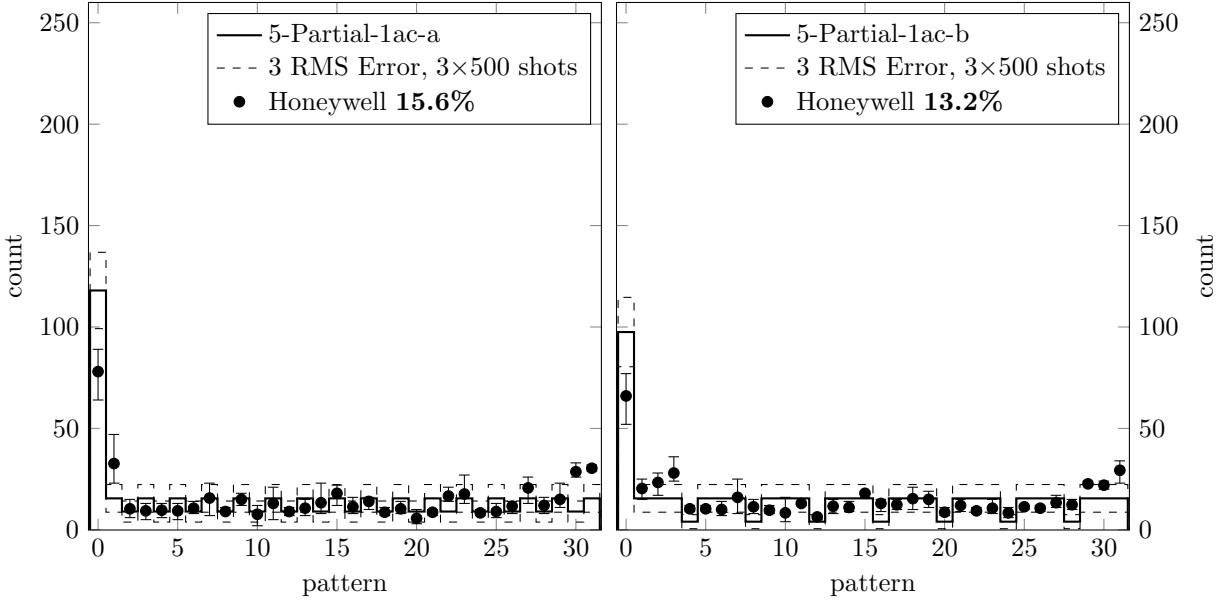
\caption{5-qubit single-iteration search utilizing single 4-qubit diffuser (left) and single 3-qubit diffuser (right)}
\end{ffigure}

\begin{ffigure}
\input plot_s2_Grover-1am-a2_5q.tex
\begin{tikzpicture}
\begin{axis}[
    height=8.6cm,  
    width=8.4cm,
    xmin=-0.6,xmax=31.6,ymin=-0.1,ymax=260,
    legend cell align=left,
    ylabel ={count}, ylabel near ticks, yticklabel pos=right,
    xlabel={pattern},
    xlabel near ticks, xticklabel pos=bottom,
]
\addplot [thick, mark=none] coordinates {
   (-0.5,0)
   (-0.5,165.0)
   (0.5,165.0)
   (0.5,48.0)
   (1.5,48.0)
   (1.5,48.0)
   (2.5,48.0)
   (2.5,48.0)
   (3.5,48.0)
   (3.5,1.0)
   (4.5,1.0)
   (4.5,9.0)
   (5.5,9.0)
   (5.5,9.0)
   (6.5,9.0)
   (6.5,9.0)
   (7.5,9.0)
   (7.5,1.0)
   (8.5,1.0)
   (8.5,9.0)
   (9.5,9.0)
   (9.5,9.0)
   (10.5,9.0)
   (10.5,9.0)
   (11.5,9.0)
   (11.5,1.0)
   (12.5,1.0)
   (12.5,9.0)
   (13.5,9.0)
   (13.5,9.0)
   (14.5,9.0)
   (14.5,9.0)
   (15.5,9.0)
   (15.5,1.0)
   (16.5,1.0)
   (16.5,9.0)
   (17.5,9.0)
   (17.5,9.0)
   (18.5,9.0)
   (18.5,9.0)
   (19.5,9.0)
   (19.5,1.0)
   (20.5,1.0)
   (20.5,9.0)
   (21.5,9.0)
   (21.5,9.0)
   (22.5,9.0)
   (22.5,9.0)
   (23.5,9.0)
   (23.5,1.0)
   (24.5,1.0)
   (24.5,9.0)
   (25.5,9.0)
   (25.5,9.0)
   (26.5,9.0)
   (26.5,9.0)
   (27.5,9.0)
   (27.5,1.0)
   (28.5,1.0)
   (28.5,9.0)
   (29.5,9.0)
   (29.5,9.0)
   (30.5,9.0)
   (30.5,9.0)
   (31.5,9.0)
   (31.5,0)
};
\addplot [dashed, mark=none] coordinates { 
   (-0.5,0)
   (-0.5,165.0+22.24859546128699)
   (0.5,165.0+22.24859546128699)
   (0.5,48.0+12.0)
   (1.5,48.0+12.0)
   (1.5,48.0+12.0)
   (2.5,48.0+12.0)
   (2.5,48.0+12.0)
   (3.5,48.0+12.0)
   (3.5,1.0+1.7320508075688774)
   (4.5,1.0+1.7320508075688774)
   (4.5,9.0+5.196152422706632)
   (5.5,9.0+5.196152422706632)
   (5.5,9.0+5.196152422706632)
   (6.5,9.0+5.196152422706632)
   (6.5,9.0+5.196152422706632)
   (7.5,9.0+5.196152422706632)
   (7.5,1.0+1.7320508075688774)
   (8.5,1.0+1.7320508075688774)
   (8.5,9.0+5.196152422706632)
   (9.5,9.0+5.196152422706632)
   (9.5,9.0+5.196152422706632)
   (10.5,9.0+5.196152422706632)
   (10.5,9.0+5.196152422706632)
   (11.5,9.0+5.196152422706632)
   (11.5,1.0+1.7320508075688774)
   (12.5,1.0+1.7320508075688774)
   (12.5,9.0+5.196152422706632)
   (13.5,9.0+5.196152422706632)
   (13.5,9.0+5.196152422706632)
   (14.5,9.0+5.196152422706632)
   (14.5,9.0+5.196152422706632)
   (15.5,9.0+5.196152422706632)
   (15.5,1.0+1.7320508075688774)
   (16.5,1.0+1.7320508075688774)
   (16.5,9.0+5.196152422706632)
   (17.5,9.0+5.196152422706632)
   (17.5,9.0+5.196152422706632)
   (18.5,9.0+5.196152422706632)
   (18.5,9.0+5.196152422706632)
   (19.5,9.0+5.196152422706632)
   (19.5,1.0+1.7320508075688774)
   (20.5,1.0+1.7320508075688774)
   (20.5,9.0+5.196152422706632)
   (21.5,9.0+5.196152422706632)
   (21.5,9.0+5.196152422706632)
   (22.5,9.0+5.196152422706632)
   (22.5,9.0+5.196152422706632)
   (23.5,9.0+5.196152422706632)
   (23.5,1.0+1.7320508075688774)
   (24.5,1.0+1.7320508075688774)
   (24.5,9.0+5.196152422706632)
   (25.5,9.0+5.196152422706632)
   (25.5,9.0+5.196152422706632)
   (26.5,9.0+5.196152422706632)
   (26.5,9.0+5.196152422706632)
   (27.5,9.0+5.196152422706632)
   (27.5,1.0+1.7320508075688774)
   (28.5,1.0+1.7320508075688774)
   (28.5,9.0+5.196152422706632)
   (29.5,9.0+5.196152422706632)
   (29.5,9.0+5.196152422706632)
   (30.5,9.0+5.196152422706632)
   (30.5,9.0+5.196152422706632)
   (31.5,9.0+5.196152422706632)
   (31.5,0)
};
\addplot [only marks, error bars,y dir=both, y explicit] coordinates {
(0, 69.9) += (0, 50.85) -= (0, 44.900000000000006)
(1, 23.650000000000002) += (1, 5.35) -= (1, 9.65)
(2, 26.3) += (2, 3.7) -= (2, 6.3)
(3, 24.0) += (3, 7.95) -= (3, 6.0)
(4, 15.350000000000001) += (4, 7.6499999999999995) -= (4, 7.35)
(5, 16.0) += (5, 10.0) -= (5, 7.0)
(6, 11.350000000000001) += (6, 6.6499999999999995) -= (6, 4.35)
(7, 11.299999999999999) += (7, 3.7) -= (7, 5.3)
(8, 16.0) += (8, 13.0) -= (8, 10.0)
(9, 11.65) += (9, 6.35) -= (9, 7.6499999999999995)
(10, 12.65) += (10, 5.35) -= (10, 8.65)
(11, 8.35) += (11, 1.65) -= (11, 1.35)
(12, 9.350000000000001) += (12, 8.65) -= (12, 6.35)
(13, 6.35) += (13, 2.65) -= (13, 4.35)
(14, 9.0) += (14, 8.0) -= (14, 7.0)
(15, 9.350000000000001) += (15, 3.65) -= (15, 4.35)
(16, 18.0) += (16, 6.0) -= (16, 5.0)
(17, 14.65) += (17, 3.35) -= (17, 3.7)
(18, 14.0) += (18, 4.0) -= (18, 6.0)
(19, 11.350000000000001) += (19, 2.65) -= (19, 2.35)
(20, 13.65) += (20, 14.35) -= (20, 7.6499999999999995)
(21, 15.0) += (21, 11.0) -= (21, 10.0)
(22, 11.350000000000001) += (22, 3.65) -= (22, 2.35)
(23, 10.0) += (23, 7.0) -= (23, 4.0)
(24, 6.6499999999999995) += (24, 2.35) -= (24, 2.65)
(25, 12.0) += (25, 7.0) -= (25, 9.0)
(26, 10.3) += (26, 3.65) -= (26, 6.3)
(27, 9.0) += (27, 6.0) -= (27, 6.0)
(28, 17.0) += (28, 24.0) -= (28, 12.0)
(29, 18.0) += (29, 9.0) -= (29, 6.0)
(30, 21.0) += (30, 24.0) -= (30, 18.0)
(31, 17.65) += (31, 13.350000000000001) -= (31, 8.65)
};
\addplot [dashed, mark=none] coordinates {
   (-0.5,0)
   (-0.5,165.0-22.24859546128699)
   (0.5,165.0-22.24859546128699)
   (0.5,48.0-12.0)
   (1.5,48.0-12.0)
   (1.5,48.0-12.0)
   (2.5,48.0-12.0)
   (2.5,48.0-12.0)
   (3.5,48.0-12.0)
   (3.5,1.0-1.7320508075688774)
   (4.5,1.0-1.7320508075688774)
   (4.5,9.0-5.196152422706632)
   (5.5,9.0-5.196152422706632)
   (5.5,9.0-5.196152422706632)
   (6.5,9.0-5.196152422706632)
   (6.5,9.0-5.196152422706632)
   (7.5,9.0-5.196152422706632)
   (7.5,1.0-1.7320508075688774)
   (8.5,1.0-1.7320508075688774)
   (8.5,9.0-5.196152422706632)
   (9.5,9.0-5.196152422706632)
   (9.5,9.0-5.196152422706632)
   (10.5,9.0-5.196152422706632)
   (10.5,9.0-5.196152422706632)
   (11.5,9.0-5.196152422706632)
   (11.5,1.0-1.7320508075688774)
   (12.5,1.0-1.7320508075688774)
   (12.5,9.0-5.196152422706632)
   (13.5,9.0-5.196152422706632)
   (13.5,9.0-5.196152422706632)
   (14.5,9.0-5.196152422706632)
   (14.5,9.0-5.196152422706632)
   (15.5,9.0-5.196152422706632)
   (15.5,1.0-1.7320508075688774)
   (16.5,1.0-1.7320508075688774)
   (16.5,9.0-5.196152422706632)
   (17.5,9.0-5.196152422706632)
   (17.5,9.0-5.196152422706632)
   (18.5,9.0-5.196152422706632)
   (18.5,9.0-5.196152422706632)
   (19.5,9.0-5.196152422706632)
   (19.5,1.0-1.7320508075688774)
   (20.5,1.0-1.7320508075688774)
   (20.5,9.0-5.196152422706632)
   (21.5,9.0-5.196152422706632)
   (21.5,9.0-5.196152422706632)
   (22.5,9.0-5.196152422706632)
   (22.5,9.0-5.196152422706632)
   (23.5,9.0-5.196152422706632)
   (23.5,1.0-1.7320508075688774)
   (24.5,1.0-1.7320508075688774)
   (24.5,9.0-5.196152422706632)
   (25.5,9.0-5.196152422706632)
   (25.5,9.0-5.196152422706632)
   (26.5,9.0-5.196152422706632)
   (26.5,9.0-5.196152422706632)
   (27.5,9.0-5.196152422706632)
   (27.5,1.0-1.7320508075688774)
   (28.5,1.0-1.7320508075688774)
   (28.5,9.0-5.196152422706632)
   (29.5,9.0-5.196152422706632)
   (29.5,9.0-5.196152422706632)
   (30.5,9.0-5.196152422706632)
   (30.5,9.0-5.196152422706632)
   (31.5,9.0-5.196152422706632)
   (31.5,0)
};
\addlegendentry{5-PDrzewker-1ac-pu}
\addlegendentry {3 RMS Error, 3$\times$500 shots}
\addlegendentry{Honeywell {\bf 14.0\%}} 
\end{axis}
\end{tikzpicture}
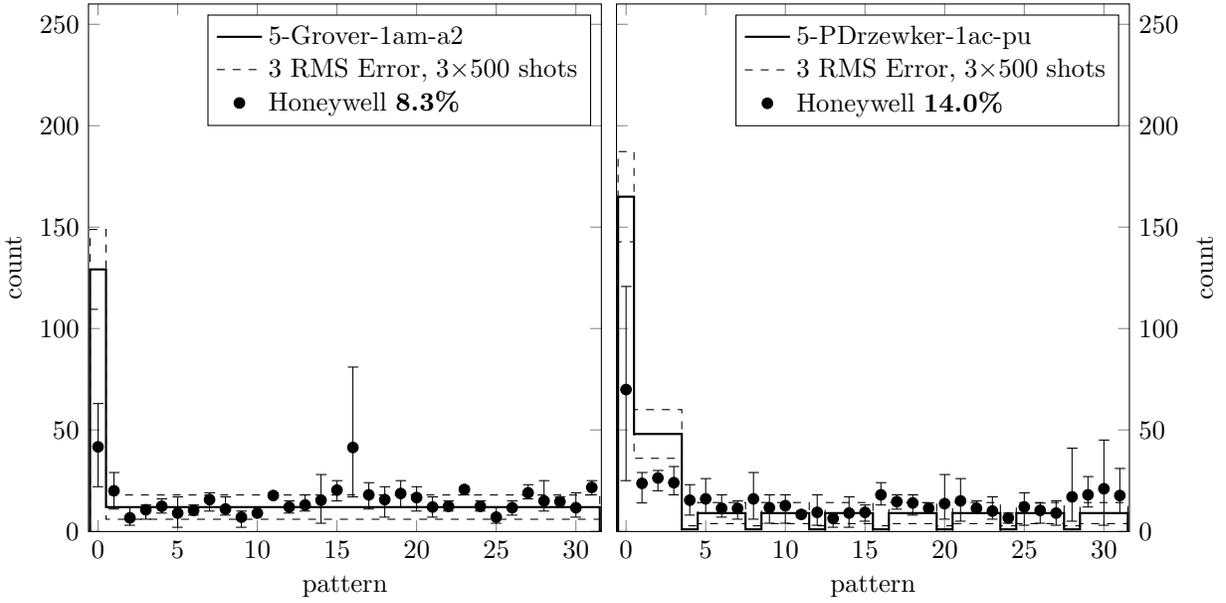
\caption{5-qubit Grover with measurement-enhanced C*Z gates (left).
\\
Partial Drzewker, only performing the first two diffusion operators (right).}
\end{ffigure}

\pagebreak
\subsection{6-qubit search}

\begin{ffigure}
\input plot_s1_Partial-a_6q.tex
\input plot_s2_Partial-b_6q.tex
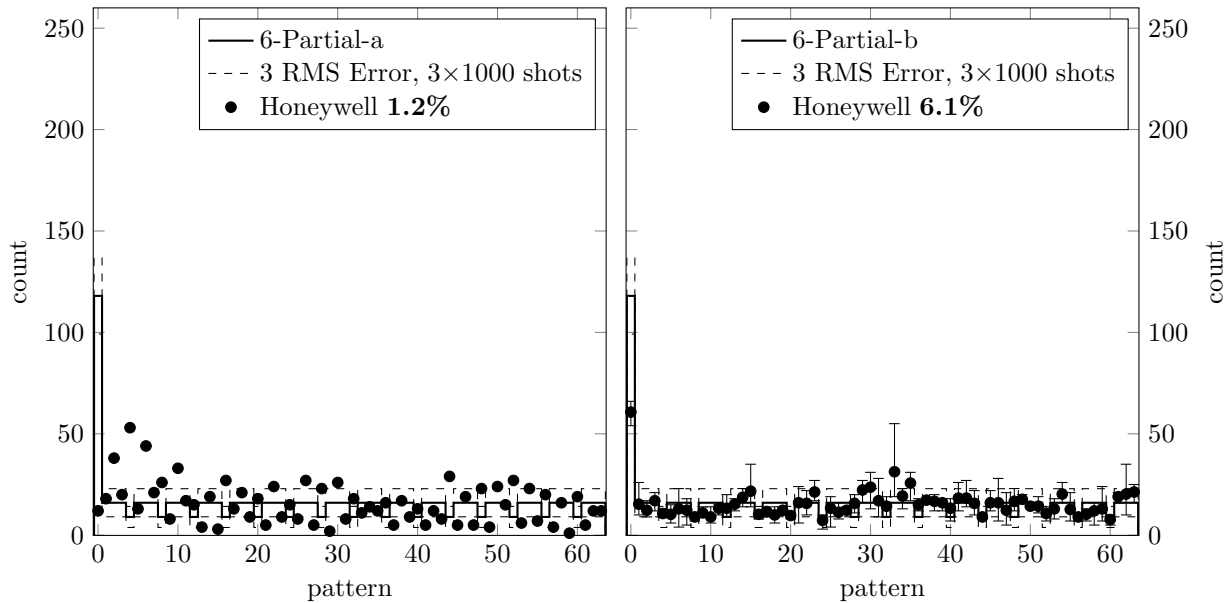
\caption{6-qubit single-iteration search utilizing single 3-qubit diffuser from session 1 (left), single 3-qubit diffuser and oracle implemented by Toffoli phase relative gate from session 2 (right).}
\end{ffigure}


\section{Conclusions and further work}\label{sec:conclusions}

The experiments performed show a significant improvement in probability of success over previous attempts.

One can observe that on Honeywell System Model H0 hardware it is sufficient to apply (heavily optimized) out-of-the-box Grover algorithm for 3-qubit search.  However, using different bespoke ideas for the larger number of qubits results in much higher probability of success. 
To our best
knowledge, the approach presented is the first 6-qubit search demonstrating the probability of success of unstructured search higher than classically possible. The specific techniques proposed by authors delivered improvements over previous results in smaller spaces where they existed, and achieved the results limited only by the qubit count currently available on the Honeywell System Model H0 platform.

For these spaces, circuits taking advantage of mid-circuit-measurement do not beat traditional deferred measurement approach yet. However, they constitute a proof that this approach works in practice. Our current work focuses on improving them.

\section*{Acknowledgements}
We would like to thank the Honeywell Quantum Solutions team 
for excellent support and collaboration during our runs.

\bibliographystyle{alpha}
\addcontentsline{toc}{section}{References}
\bibliography{references}

\newcommand{\etalchar}[1]{$^{#1}$}
\begin{thebibliography}{DHHM06}

\bibitem[BGH{\etalchar{+}}20]{prackum}
Marcin Briański, Jan Gwinner, Vladyslav Hlembotskyi, Witold Jarnicki, Szymon
  Pliś, and Adam Szady.
\newblock Introducing structure to expedite quantum search, 2020.

\bibitem[BHT98]{Brassard_1998}
Gilles Brassard, Peter HØyer, and Alain Tapp.
\newblock Quantum cryptanalysis of hash and claw-free functions.
\newblock {\em Lecture Notes in Computer Science}, page 163–169, 1998.

\bibitem[DHHM06]{D_rr_2006}
Christoph Dürr, Mark Heiligman, Peter HOyer, and Mehdi Mhalla.
\newblock Quantum query complexity of some graph problems.
\newblock {\em SIAM Journal on Computing}, 35(6):1310–1328, Jan 2006.

\bibitem[FML{\etalchar{+}}17]{3q}
C.~Figgatt, D.~Maslov, K.~Landsman, N.~Linke, Shantanu Debnath, and C.~Monroe.
\newblock Complete 3-qubit grover search on a programmable quantum computer.
\newblock {\em Nature Communications}, 8, 03 2017.

\bibitem[GBB{\etalchar{+}}20]{prackum2}
Jan Gwinner, Marcin Briański, Wojciech Burkot, Łukasz Czerwiński, and
  Vladyslav Hlembotskyi.
\newblock Benchmarking 16-element quantum search algorithms on ibm quantum
  processors, 2020.

\bibitem[Gro96]{grover96}
Lov~K Grover.
\newblock A fast quantum mechanical algorithm for database search.
\newblock In {\em Proceedings of the twenty-eighth annual ACM symposium on
  Theory of computing}, pages 212--219, 1996.

\bibitem[KS18]{Stromberg}
VERA~BLOMKVIST KARLSSON and PHILIP STR{\"O}MBERG.
\newblock 4-qubit grover's algorithm implemented for the ibmqx5 architecture.
\newblock 2018.

\bibitem[Mas16]{adamolus}
Dmitri Maslov.
\newblock Advantages of using relative-phase toffoli gates with an application
  to multiple control toffoli optimization.
\newblock {\em Physical Review A}, 93(2), Feb 2016.

\bibitem[MOJ18]{Mandviwalla}
A.~{Mandviwalla}, K.~{Ohshiro}, and B.~{Ji}.
\newblock Implementing grover’s algorithm on the ibm quantum computers.
\newblock In {\em 2018 IEEE International Conference on Big Data (Big Data)},
  pages 2531--2537, 2018.

\bibitem[SOM20]{satoh2020subdivided}
Takahiko Satoh, Yasuhiro Ohkura, and Rodney~Van Meter.
\newblock Subdivided phase oracle for nisq search algorithms, 2020.

\end{thebibliography}

\end{document}